\documentclass[manuscript]{aastex}

\usepackage{subfigure}
\usepackage{natbib}
\usepackage{graphicx}
\usepackage{multirow}
\usepackage{epstopdf}
\usepackage{amsmath}
\usepackage{url}
\usepackage{hyperref}
\usepackage{amssymb}
\usepackage{booktabs}
\usepackage{lineno}
\usepackage{color}

\newcommand{\Swift}{{\it Swift~}}

\shorttitle{GRB Energetics}
\shortauthors{Goldstein, et al.}

\begin{document}

\title{Estimating Long GRB Jet Opening Angles and Rest-Frame Energetics}

\author{Adam Goldstein\altaffilmark{1}, Valerie Connaughton\altaffilmark{2}, Michael S. Briggs\altaffilmark{3}, Eric Burns
\altaffilmark{3}}
\altaffiltext{1}{Space Science Office, VP62, NASA/Marshall Space Flight Center, Huntsville, AL 35812, USA}
\altaffiltext{2}{Science and Technology Institute, Universities Space Research Association, Huntsville, AL 35805, USA}
\altaffiltext{3}{Center for Space Plasma and Aeronomic Research, University of Alabama in Huntsville, 320 Sparkman Drive, 
Huntsville, AL 35899, USA}

\begin{abstract}
We present a method to estimate the jet opening angles of long duration Gamma-Ray Bursts (GRBs) using the prompt 
gamma-ray energetics and an inversion of the Ghirlanda relation, which is a correlation between the time-integrated peak 
energy of the GRB prompt spectrum and the collimation-corrected energy in gamma rays.  The derived jet opening angles 
using this method and detailed assumptions match well with the corresponding inferred jet opening angles obtained when 
a break in the afterglow is observed.  Furthermore, using a model of the predicted long GRB redshift probability 
distribution observable by the {\it Fermi} Gamma-ray Burst Monitor (GBM), we estimate the probability distributions for the jet 
opening angle and rest-frame energetics for a large sample of GBM GRBs for which the redshifts have not been observed.  
Previous studies have only used a handful of GRBs to estimate these properties due to the paucity of observed afterglow jet 
breaks, spectroscopic redshifts,  and comprehensive prompt gamma-ray observations, and we potentially expand the 
number of GRBs that can be used in this analysis by more than an order of magnitude.  In this analysis, we also present an 
inferred distribution of jet breaks which indicates that a large fraction of jet breaks are not observable with current 
instrumentation and observing strategies.  We present simple parameterizations for the jet angle, energetics, and jet break 
distributions so that they may be used in future studies.
\end{abstract}

\keywords{gamma rays: bursts --- methods: data analysis}

\section{Introduction}
A key to understanding the progenitors and central engines of Gamma-Ray Bursts (GRBs) is to know the total energy budget of 
these enormous stellar explosions.  One way to estimate the total kinetic energy in a GRB is to calculate the amount of energy 
radiated in gamma rays and estimate the efficiency of converting the energy in the mass outflow of the explosion to the 
radiated energy that is observed~\citep{Frail01, Freedman01, Ghisellini02}.  Several factors affect the apparent radiated 
energy such as the physics of the mass-radiation conversion and the Lorentz factor of the relativistic jet~\citep{Kumar00}.  
These properties are difficult to estimate and are not observed directly.  In most cases even the radiated energy of a GRB is not 
readily estimated, since it requires a broadband gamma-ray modeling of the prompt emission, a set of comprehensive 
broadband observations of the afterglow to estimate the amount of jet collimation~\citep{Sari99, Frail01}, and optical 
identification of the redshift~\citep{Bloom98}.  The myriad of requisite observations and inferences to estimate the radiated 
energy in gamma rays has provided robust energetics estimates for only a few tens of GRBs compared to the several 
thousand GRBs that have been detected. 

One particular physical property of GRBs that has a large impact on the observed energetics is the degree to which the jetted 
outflow is collimated.  The amount of collimation in a particular GRB can adjust the inferred rest-frame energy 
or luminosity several orders of magnitude from an assumed isotropic explosion.  For this reason, the jet opening angle 
of GRBs is an important property to measure if inferences are to be made about their rest-frame energetics.  Unfortunately, the 
jet opening angle is difficult to reliably estimate, since it requires observations of an achromatic jet break in the power-law 
decay of the afterglow emission, most often observed in the optical and X-ray bands~\citep{Sari99, Harrison99, 
O'Brien06}.  In all but a small number of cases these observations are complicated by limited observations of the afterglow, 
rapidly fading afterglow, and late-time X-ray flaring variability in the afterglow~\citep{Costa99, O'Brien06}.  In addition to the jet 
break time, observations of the host environment and detailed afterglow spectroscopy are generally needed to estimate the 
particle density profile of the surrounding circumburst medium to estimate the jet opening angle~\citep{Waxman97, Wijers99}.  
Because of these difficulties, there are currently only $\sim$50 GRBs with reasonably constrained jet breaks and $\sim$20 of 
those GRBs have reasonable constraints on the circumburst density profile.  An additional complication is that many of the 
GRBs with constrained jet breaks do not have the broad prompt spectral coverage necessary to adequately calculate the flux 
and fluence as measured in gamma rays.

Using the available small samples of GRBs with adequate observations, a number of observed correlations between GRB 
spectral or temporal observables and the rest-frame energetics of the explosion have been discovered~\citep{Norris00, 
Amati02, Ghirlanda04, Yonetoku04, Guidorzi06}, and they have been used to investigate the physics of the prompt emission of 
GRBs.  Some of these correlations have large dispersion, in many cases too large for meaningful physical inference from the 
correlation.  One particularly tight correlation is between the rest-frame peak energy of the time-integrated 
prompt GRB spectrum and the collimation-corrected rest-frame energy in gamma rays, known as the Ghirlanda 
relation~\citep{Ghirlanda04}.  We endeavor to empirically estimate the jet opening angle for GRBs from the prolifically 
observed prompt gamma-ray emission by inverting the Ghirlanda relation.  As previous works have 
shown~\citep{Ghirlanda05a,  Ghirlanda13} this method can alleviate the problems associated with the scarcity and difficulty of 
obtaining simultaneous measurements over several energy regimes.

As we will show, the inversion of the Ghirlanda relation to estimate the jet opening angle requires knowledge 
of the GRB redshift, but only a small fraction ($<$10\%) of observed GRBs have an observed redshift.  Current observations of 
redshift are biased toward GRBs that are observed and localized by {\it Swift}, which has a gamma-ray bandpass that is 
typically too low to capture the curvature in the GRB spectrum, thereby biasing the estimate of the rest-frame gamma-ray 
energetics of the GRB.  Because of the very small fraction ($\approx$1\%) of all GRBs with both known redshift and 
broadband gamma-ray observations, there have been multiple investigations to use observed spectral and temporal 
correlations to predict the redshift~\citep{Atteia03, Yonetoku04, Ghirlanda05a, Amati06}.  The accuracy of these methods 
are difficult to assess and are fraught with systematic biases and large uncertainties~\citep{Goldstein12a}.  We propose to 
avoid this complication by using the estimated redshift probability distribution of long GRBs that is observable by a particular 
instrument--in this case, the {\it Fermi} Gamma-ray Burst Monitor (GBM).  The redshift distribution is applied as our prior 
knowledge when the redshift has not been directly observed.  This allows us to estimate the probability density functions 
(PDFs) for the rest-frame energetics of individual GRBs without known redshift as well as estimates on the population 
distributions of energetics.

\section{Methodology}

\subsection{The Ghirlanda Correlation \& Jet Opening Angles}
Estimates of the energy and luminosity require an implicit assumption about the cosmological expansion of the universe.  To 
avoid the uncertainty in assuming a cosmological model, we follow the procedure of~\citet{Liang08} by using the correlation 
between the redshift and distance (known as the Hubble Diagram) for Type Ia Supernovae (SNe Ia) to directly estimate the 
distance to GRBs with $z < 1.5$.  Since the distance of SNe Ia are well-estimated by standard lightcurve-fitting techniques, 
they provide a calibration sample for GRBs that overlap in redshift.  In general, the SNe Ia Hubble Diagram can be interpolated 
to find the distance for a given redshift.  The uncertainty in the interpolation can then be propagated to represent the 
uncertainty in the distance of the GRB.  This procedure produces a model-independent estimation of the luminosity distance 
for GRBs to be used in the calculation of the source energetics.  

Using this method to estimate the luminosity distance, we collect a sample of GRBs with $z< 1.5$ that 
have observed and published jet break time estimates.  The jet break time, redshift, and spectral properties used for these 
bursts are included in Table~\ref{GRBLoZ}.  Most GRBs in the table are fit with the traditional empirical Band 
function~\citep{Band93}, with the GRBs for which there is no high-energy index, $\beta$, in the table modeled with an 
exponentially cut-off power law.  Using the fluence and redshift of each GRB, the isotropic energy, $E_{\rm iso}$ can be 
calculated.  Because spectral and fluence measurements of the sample are from different instruments with different 
bandpasses, and due to the cosmological shifting of the spectral bandpass into the rest-frame of the GRB, the spectral 
parameters in Table~\ref{GRBLoZ} are used to calculate the cosmological K-correction as prescribed in~\citet{Bloom01}.  The 
isotropic energy is then calculated as
\begin{equation}
	E_{\rm iso} = \frac{4\pi d_L^2 S_\gamma K(z;F(E))}{1+z},
	\label{eq:Eiso}
\end{equation}
where $d_L$ is the luminosity distance, $S_\gamma$ is the fluence in the detector band, and $K(z; F(E))$ is the K-correction 
used to scale the fluence from the observed detector band to some consistent rest-frame band.  In this paper, we calculate the 
K-correction for a rest-frame energy band of 1 keV--10 MeV.

Under the assumption of the standard afterglow model and a simple uniform jet, the jet opening angle, $\theta_j$, can 
then be estimated from the measured jet break time, $t_j$, as
\begin{equation}
	\theta_j \approx 0.057 \biggl( \frac{t_j}{1\ {\rm day}} \biggr)^{3/8} \biggl( \frac{1+z}{2} \biggr)^{-3/8} \biggl( \frac{E_{\rm iso}} 
	{10^{53}\ {\rm erg} } \biggr)^{-1/8} \biggl( \frac{\epsilon}{0.2} \biggr)^{1/8} \biggl( \frac{n_p}{0.1\ {\rm cm^{-3}}} \biggr)^{1/8},
\label{eq:JetAngle}
\end{equation}
where $\epsilon$ is an estimate of the efficiency in converting the bulk kinetic outflow into gamma rays, and $n_p$ is the 
circumburst density~\citep{Sari99, Frail01}.  Both $\epsilon$ and $n_p$ are largely unknown, although some rare 
measurements of $n_p$ have been made, and $\epsilon$ is typically assumed to be 20\%.  In this work, we 
assume that the $\epsilon$ and $n_p$ are not specifically known for each GRB.  For $\epsilon$, we assume a broad uniform 
distribution spanning 5\%--95\%, consistent with estimates of GRB radiative efficiencies that span from $<10\%$ to $>90\%$~
\citep{Zhang07}.  We assume a log-normal distribution for $n_p$ with mean $\log_{10}(0.1)$ and standard deviation 1.0, 
which is derived from the distribution of the small number of measured $n_p$.  As can be seen from 
Equation~\ref{eq:JetAngle}, $\theta_j$ is less dependent on $\epsilon$ and $n_p$ than on the jet break time. Since the 
parameter uncertainties may not be strictly Gaussian, we calculate the uncertainty in $\theta_j$ by Monte Carlo sampling from 
the respective parameter probability density functions and repeatedly use Equation~\ref{eq:JetAngle} to build up the 
probability distribution for $\theta_j$.

\begin{figure}
	\begin{center}
		\includegraphics[scale=1.0]{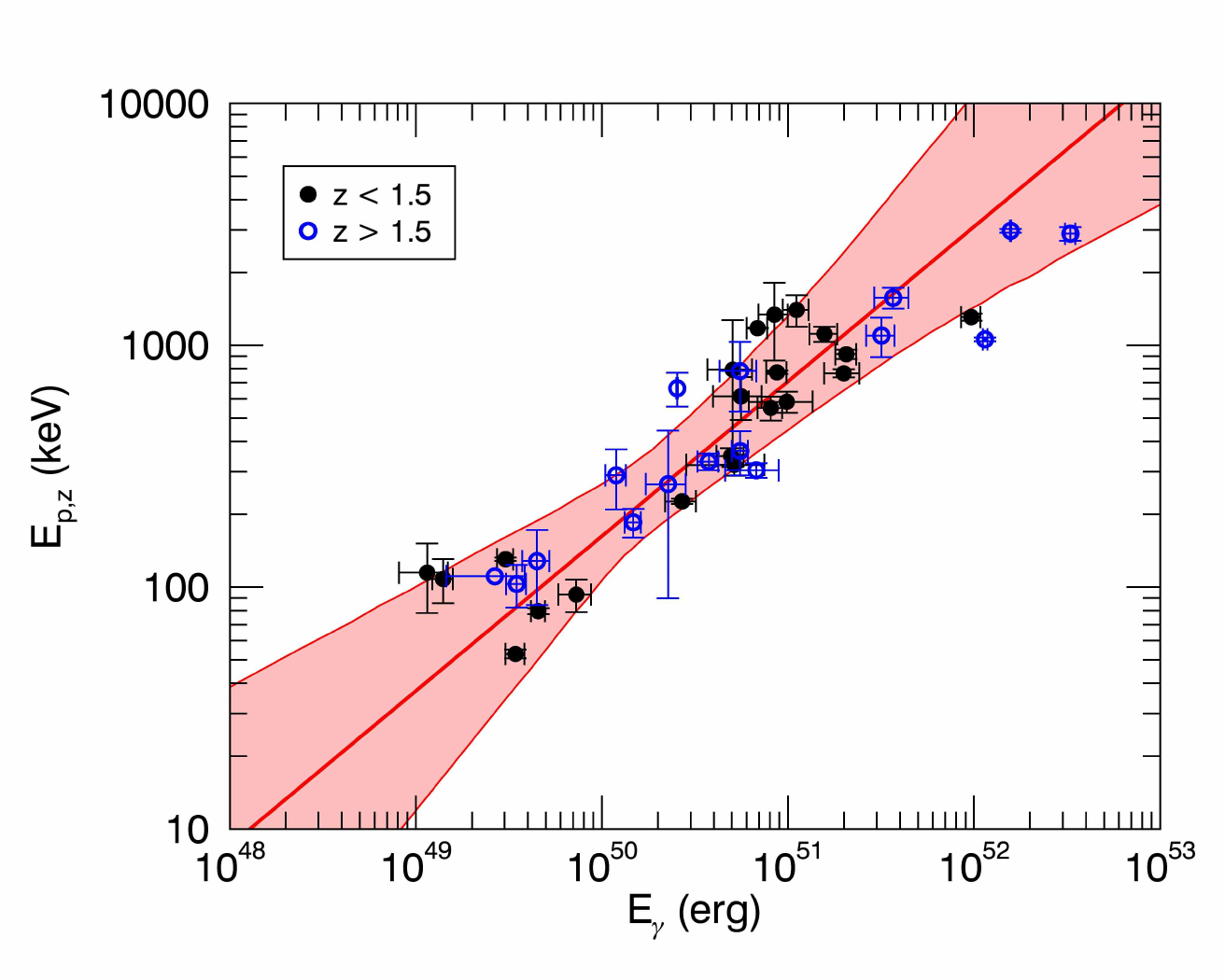}
	\end{center}
\caption{The Ghirlanda correlation between time-integrated $E_{\rm peak}$ and the peak collimation-corrected energy in 
gamma rays with 1$\sigma$ error bars.  The black filled circles are the GRBs at $z<1.5$ which are calibrated using SNe Ia, 
and the blue open circles are GRBs at $z>1.5$ which are produced assuming the concordant cosmology.  Only the low-$z$ 
data were used to fit the power law, and the red region is the 99\% credible region for the power law fit.
\label{EpzEgamma}}
\end{figure}

Having calculated $\theta_j$, the collimation-corrected energy is defined by
\begin{equation}
	E_\gamma = E_{\rm iso} [1-\cos(\theta_j)].
	\label{eq:Egamma}
\end{equation}
The peak of the $\nu F_\nu$ spectrum, known as $E_{\rm peak}$, in the rest-frame, $E_{p,z}$, can then be plotted against 
$E_\gamma$, which shows the observed 
Ghirlanda relation, displayed in Figure~\ref{EpzEgamma}.  The estimated best fit for the power law using a Bayesian method taking into account uncertainties in both variables and intrinsic scatter~\citep{Kelly07} is:
\begin{equation}
	\frac{E_{p,z}}{1 \ \rm keV} = 10^{2.57\pm0.08} \biggl( \frac{E_\gamma}{3.8\times10^{50} \ \rm erg} \biggr)^{0.61\pm0.10}.
	\label{eq:EpzEgammaFit}
\end{equation}
The correlation index is in good agreement with the 2/3 prediction of~\citet{Levinson05}, in which the $E_{p,z}-E_\gamma$ 
correlation was derived using jet dynamics and off-axis viewing effects of a simple annular jet and is consistent with 
previous findings~\citep{Ghirlanda07}.  Figure~\ref{EpzEgamma} also displays the $E_{p,z}-E_\gamma$ correlation for a 
sample of GRBs at high redshift ($z > 1.5$).  The correlation at higher redshift is generally consistent with the the calibrated 
correlation and the parameters used for these GRBs can be found in Table~\ref{GRBHiZ}.  When using values of redshift at 
$z>1.5$, we assume the concordant $\Lambda$CDM cosmology with $\Omega_M = 0.27$, $\Omega_\Lambda = 0.73$, and 
$H_0 = 70 \ \rm km \ Mpc^{-1} \ s^{-1}$.

Using Equations~\ref{eq:Eiso},~\ref{eq:Egamma},~and~\ref{eq:EpzEgammaFit}, the correlation can be inverted to estimate $
\theta_j$ given the time-integrated observed spectrum, fluence, and redshift:
\begin{equation}
	\theta_j  = \cos^{-1}\Biggl(1- \frac{3.8\times10^{50}}{4\pi} \frac{1+z}{d_L^2 S_\gamma K} \biggl[\frac{E_{\rm peak} (1+z)}
	 {\xi}\biggr]^{1/\eta} \Biggr),
	\label{eq:predictTheta}
\end{equation}
where $\xi = 10^{2.57\pm0.08}$ and $\eta=0.61\pm0.10$ are the correlation amplitude and index, respectively.
We choose to calculate the uncertainties on $\theta_j$ via Monte Carlo sampling of $d_L$, $S_\gamma$, $K$, $E_{peak}$, $
\xi$, and $\eta$.  We do this because the PDFs for each of these quantities are not necessarily Gaussian or even symmetric, 
and this method is further required when the redshift is not explicitly known, as discussed in the next section.

\subsection{Redshift Distribution of GRBs Observable by GBM}
The GBM-observable GRB redshift distribution can be estimated by taking into account the detector sensitivity, the 
detector-dependent GRB luminosity function, and the GRB rate density evolution.  We follow this method detailed 
in~\citet{Coward13}, who used the method to produce the redshift distribution for {\it Swift} GRBs.  Specifically, 
the GRB redshift distribution observable by GBM can be written as 
\begin{equation}
	P(z) = N_p \frac{dV(z)}{dz} \frac{e(z)}{1+z} \psi_{GBM}(z),
	\label{eq:zDistrib}
\end{equation}
where $dV(z)/dz$ is the comoving cosmological volume element, $e(z)$ is the GRB rate evolution model, $\psi_{GBM}(z)$ is 
the GBM sensitivity to detecting GRBs at redshift $z$, and $N_p$ is the normalization.  \citet{Coward13} showed that when 
a complete sample of observed {\it Swift} GRBs with redshift was studied factoring in the {\it Swift} detector biases as well as 
detailed optical biases that affect the observation of redshift, neither luminosity nor density evolution for GRBs was required to 
explain the observed GRB rate deviation from the star formation rate.  Therefore $e(z)$ represents a parametrization of the star 
formation rate that is normalized to the local GRB rate density.  $\psi_{GBM}$ is calculated by estimating the detector-
dependent luminosity function and integrating it over observable luminosities.  Following~\citet{Howell13}, we use an 
exponentially cut-off power law to model the luminosity function of GBM GRBs:
\begin{equation}
	\phi(L) = \phi_0 \biggl( \frac{L}{L_\ast} \biggr)^{\alpha} \exp \biggl(-\frac{L_\ast}{L}\biggr),
\end{equation}
where $\phi_0$ is the normalization and the best fit parameters from the differential log N--log P distribution is $L_\ast = 
\bigl(4.66^{+0.09}_{-0.48}\bigr)\times10^{52} \rm \ erg \ s^{-1}$ and $\alpha = -4.03^{+0.16}_{-0.05}$.  Note that these 
parameters are consistent with~\citet{Howell13} and the no-evolution model in~\citet{Salvaterra07} for {\it Swift} GRBs.  The 
luminosity function is integrated starting at a lower-limiting luminosity defined by the lowest peak flux observed:
\begin{equation}
	\psi_{GBM}(z) = \int_{L_{\rm lim}(F_{\rm lim}, z)}^{L_{\rm max}} \phi(L) dL,
\end{equation}
where $L_{\rm lim}$ is the limiting luminosity, which is a function of the limiting flux, 
$F_{\rm lim}$ (0.8 $\rm ph \ s^{-1} \ cm^{-2}$ in 10--1000 keV for GBM), and $z$.  The observable redshift distribution is 
shown in Figure~\ref{zDistrib} and is compared to the distribution of long GRBs with measured spectroscopic redshift through 
March 2015~\citep{Greiner15}.  The distribution of 40 GBM GRBs from the published four-year GBM Catalog~\citep{Gruber14} 
with observed redshift is also compared against the theoretical redshift distribution.  The obvious difference between the 
redshift distribution for observed GBM GRBs and the theoretical observable redshift distribution may be explained by the 
requirement that either the \Swift BAT must have observed the GRB or that the GRB was particularly bright and seen in the {\it 
Fermi} LAT.  Indeed, if Equation~\ref{eq:zDistrib} is fit to the observed GBM redshift distribution, the resulting limiting flux 
increases more than an order of magnitude to $F_{\rm lim}=10.5$, indicating that redshift observations for GBM GRBs is 
biased toward brighter bursts.  Because the energy range,  exposure, and sensitivity of instruments required to observe the 
prompt emission of a GBM burst and trigger follow-up observations are different from that of GBM, the current observed 
distribution of redshifts for GBM-detected GRBs is not guaranteed to be consistent with the true redshift distribution of all GRBs 
detected by GBM.

\subsection{Estimation of Histograms}
The jet opening angle and energetics distributions for GRBs without known redshift are constructed from probability 
distributions of the values in question that are much broader than if the individual redshift values are known.  For this reason, 
binning the distributions requires some care.  Typically histograms are produced by binning continuous data and the resulting 
bins are treated as a Poisson random variable, and therefore, for the $n_i$ items in the $i$th bin, the (1$\sigma$) uncertainty 
is modeled as $\sqrt{n_i}$.  This assumption is not appropriate for some of the distributions in this paper.  Instead, we choose 
to create histograms via a Monte Carlo sampling from the PDF of each quantity from each GRB.  Specifically, for a quantity of 
interest from $N$ total GRBs in our sample, we first determine the edges of our bins, then we take a sample from each of the 
$N$ PDFs and place them in the corresponding bins.  This is done for a number of iterations (typically $>1000$), randomly 
sampling from the PDFs and recording the counts in each bin for each iteration.   This process creates a PDF for each bin of 
the histogram, from which we choose the median as the centroid of the bin and the error bars represent the 68\% credible 
interval centered at the median.  This Monte Carlo sampling method allows us to more accurately represent the underlying 
distribution, especially at the extremes of the distribution where a combination of several low probability densities can produce 
a non-negligible probability density in the histogram.  This method is applied to the histograms presented in the following 
section.

\section{Data Analysis \& Results}
To study the rest-frame energetics we use the results from the {\it Fermi} GBM GRB spectroscopy catalog~\citep{Gruber14}, 
which covers the first four years of GBM observations.  The catalog contains both the time-integrated spectral fits and the 
spectral fits at the peak flux for each of 943 GRBs.  We only consider those GRBs which are defined as long by the centroid of 
the $T_{90}$ duration estimate, namely $T_{90} > 2$ s. To obtain a reliable estimate of $E_{\rm peak}$, we only consider long 
GRBs from that catalog which are adequately fit by the empirical Band function or an exponentially-cutoff power law, known as 
a Comptonized function.  Specifically, we first consider long GRBs from the catalog that are well fit by a Band function with 
well-constrained parameters which results in 381 GRBs (the GOOD criteria defined in \cite{Gruber14}). Of the remaining long 
GRBs, we add to our sample GRBs that are well fit by the Comptonized function as specified in the catalog, which results in an 
additional 257 GRBs.  We  use this sample of 638 long GRBs to study the $\theta_j$ and energetics distributions.

The following subsections describe the estimation of $\theta_j$ from the inversion of the Ghirlanda relation for GRBs with 
and without known redshift. This estimation of $\theta_j$, particularly for GRBs without known redshift, is then applied to 
calculate the distributions of the rest-frame energetics.  We also look at the correlations between three $E_{\rm peak}$ 
correlations: $E_{p,z}-E_{\rm iso}$, $E_{p,z}-L_{\rm iso}$, and $E_{p,z}-L_\gamma$.  Finally, we present an estimate of the jet 
break time distribution which allow us to make predictions about the likelihood of directly observing jet breaks.  Note that all of 
the PDFs that are generated are well modeled as log-normals.  We parametrize the log-normal PDFs as
$P(\log_{10}{x}) = N(\mu,\sigma)$ so that the $\mu$ and $\sigma$ values quoted are for a normal distribution in $\log_{10}$ 
space.  The best-fit parameters for the jet angle, energetics, and jet break time distributions can be found in 
Table~\ref{distributionParams}.  The log-normal parameters for all estimated quantities for each GRB in the GBM sample are 
listed in machine-readable format which is described in Table~\ref{pdfParams}. 

\subsection{Predicting Jet Opening Angles}
\begin{figure}
	\begin{center}
		\subfigure[]{\label{thetaWithZ}\includegraphics[scale=0.35]{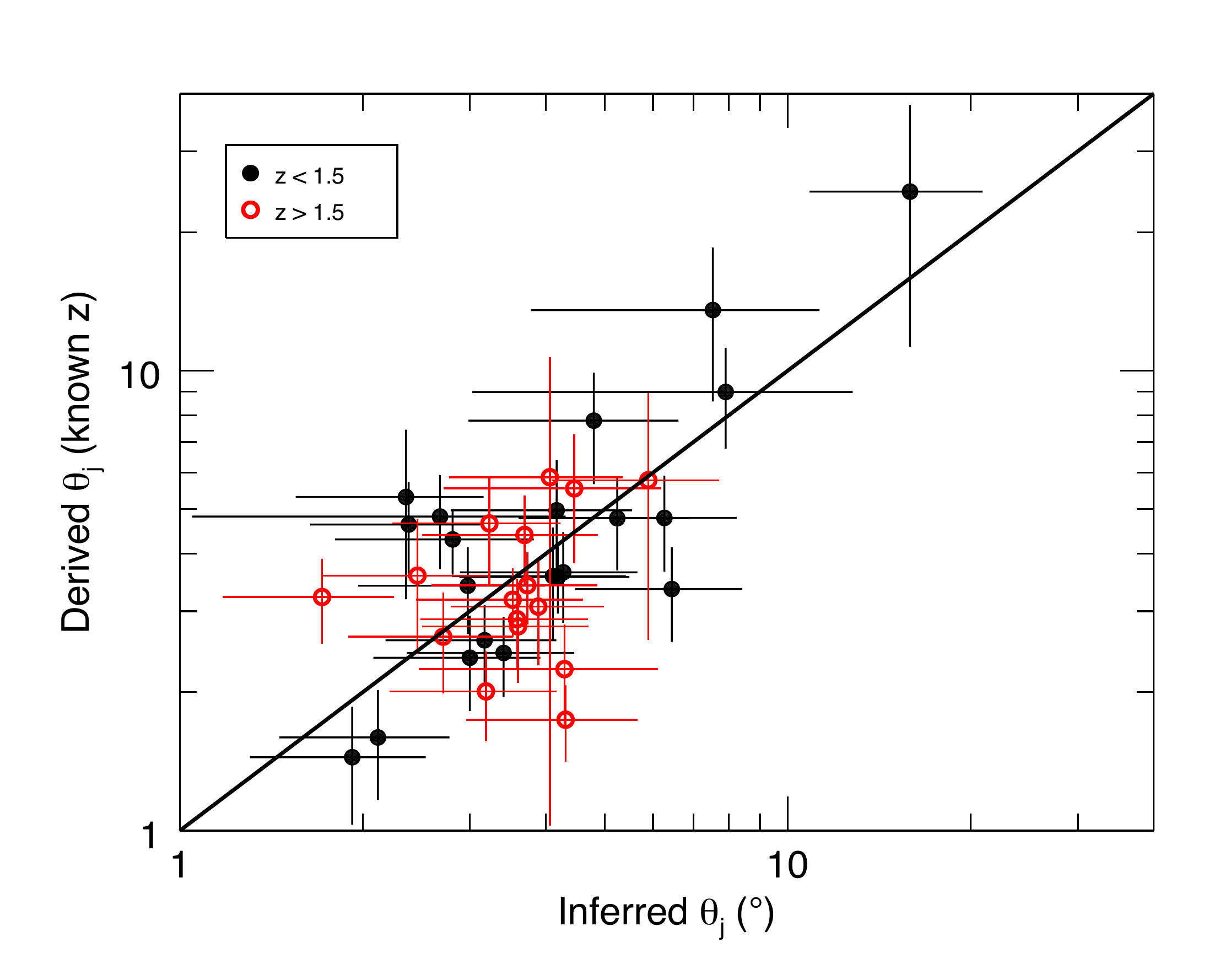}}
		\subfigure[]{\label{thetaNoZ}\includegraphics[scale=0.35]{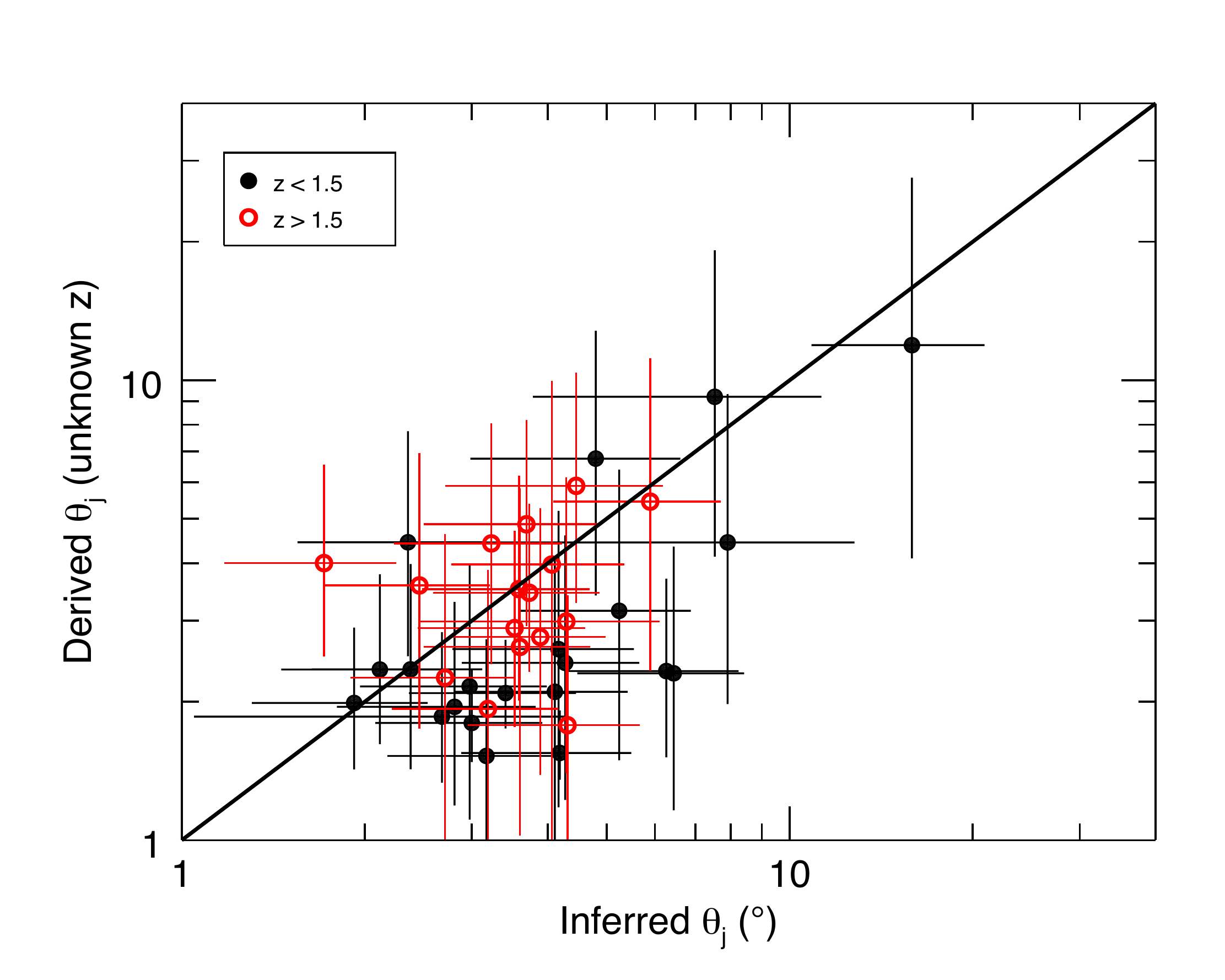}}\\
		\subfigure[]{\label{thetaZ_noZ}\includegraphics[scale=0.35]{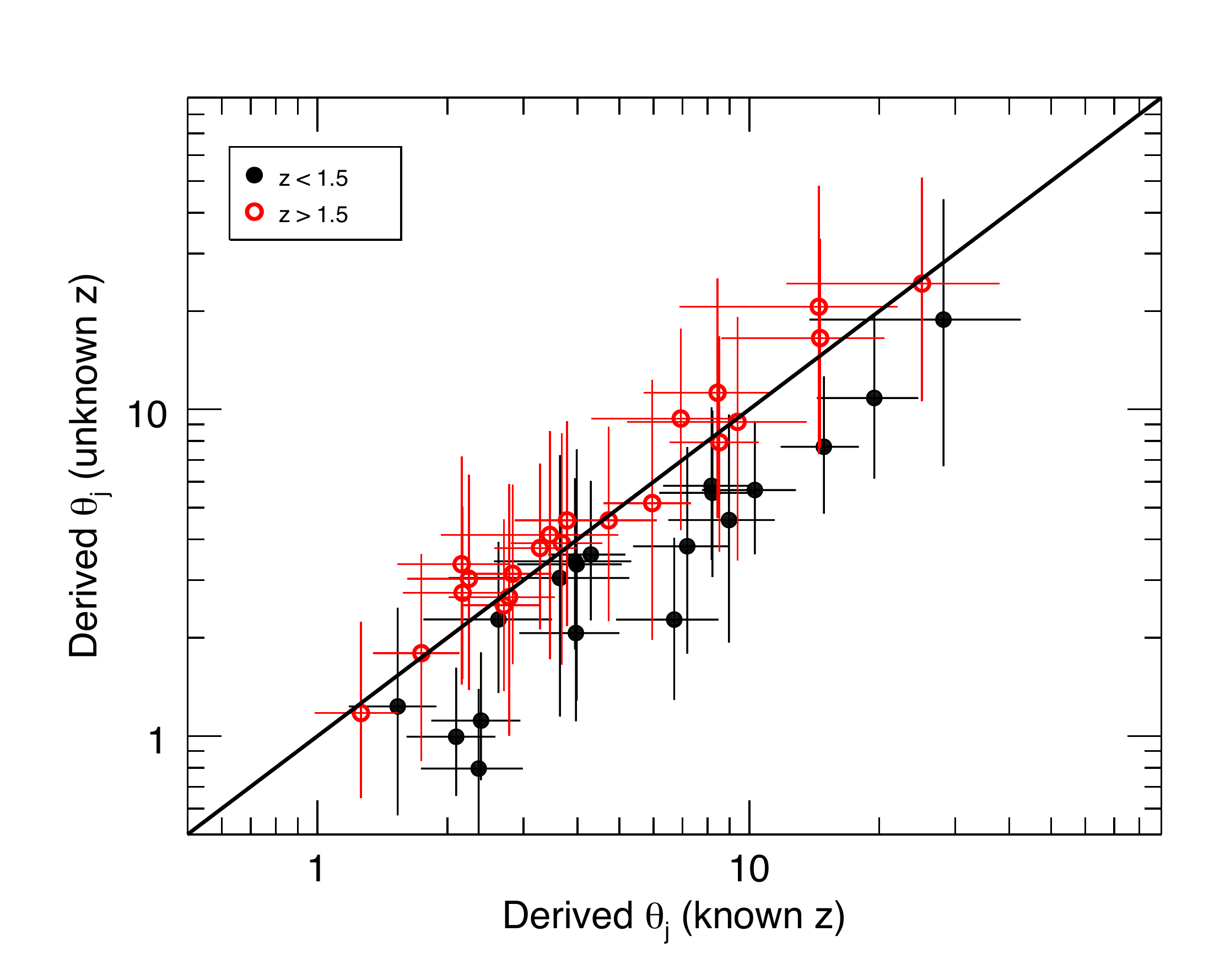}}
		\subfigure[]{\label{theta_Z}\includegraphics[scale=0.35]{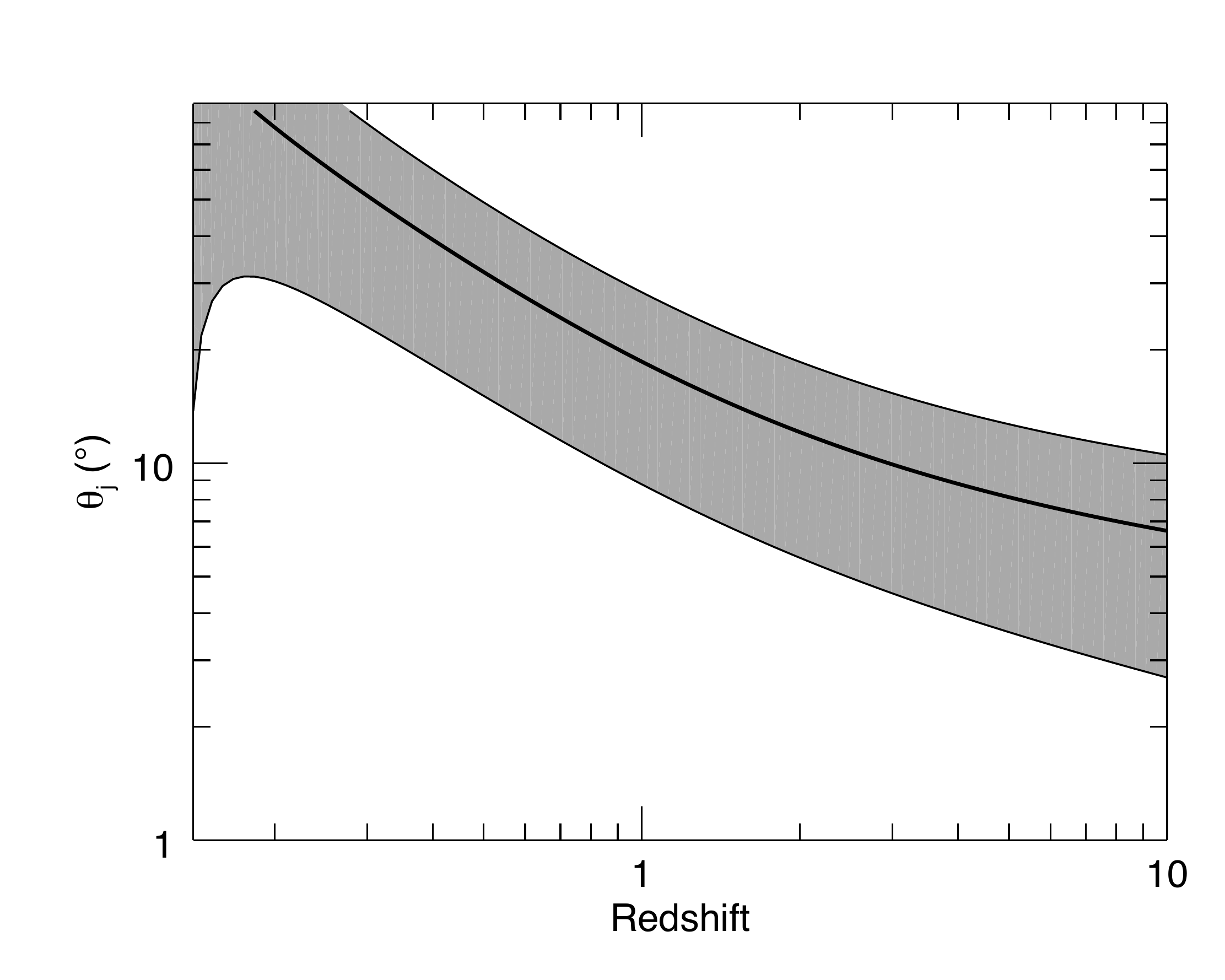}}
	\end{center}
\caption{Panel~\ref{thetaWithZ} shows a comparison of the jet opening angle inferred from the afterglow jet break and the jet 
opening angle derived from the Ghirlanda relation.  The black filled circles are the low-redshift calibration sample, and the red 
open circles are the high-redshift sample assuming $\Lambda$CDM. Panel~\ref{thetaNoZ} is a comparison of the jet opening 
angle inferred from the jet break and assuming the redshift is unknown. Panel~\ref{thetaZ_noZ} shows the estimation of $
\theta_j$ for 40 long GRBs from the GBM catalog with observed redshift.  Using the observed redshift to estimate $\theta_j$ is 
compared to assuming the GBM GRB redshift distribution.  The systematic difference between GRBs with low- and 
high-redshifts can be explained by the functional dependence of $\theta_j$ on redshift, as shown in Panel~\ref{theta_Z}, where 
higher redshift will tend to result in smaller $\theta_j$ for a fixed $E_{\rm peak}$ and fluence.
\label{JetAngleCompare}}
\end{figure}

For GRBs with observed redshift, $\theta_j$ can be estimated by Equation~\ref{eq:predictTheta}.  An important, but often 
ignored, aspect of this estimation is a proper propagation of uncertainty.  We propagate the uncertainty in the fluence, $E_{\rm 
peak}$, K-correction, luminosity distance, and the correlation parameters to estimate the uncertainty in $\theta_j$ for this 
method.  A comparison of the inferred jet opening angle using the measured jet break and the estimate of the jet opening 
angle from this paper is shown in Figure~\ref{thetaWithZ}.  We find that 84\% of the estimates for $\theta_j$ are consistent 
with the calculation of $\theta_j$ via observed jet breaks within the combined 1$\sigma$ confidence level.  Our propagation of 
uncertainty is shown to capture the uncertainty in the spectral fit and the Ghirlanda correlation. 

\begin{figure}
	\begin{center}
		\includegraphics[scale=0.50]{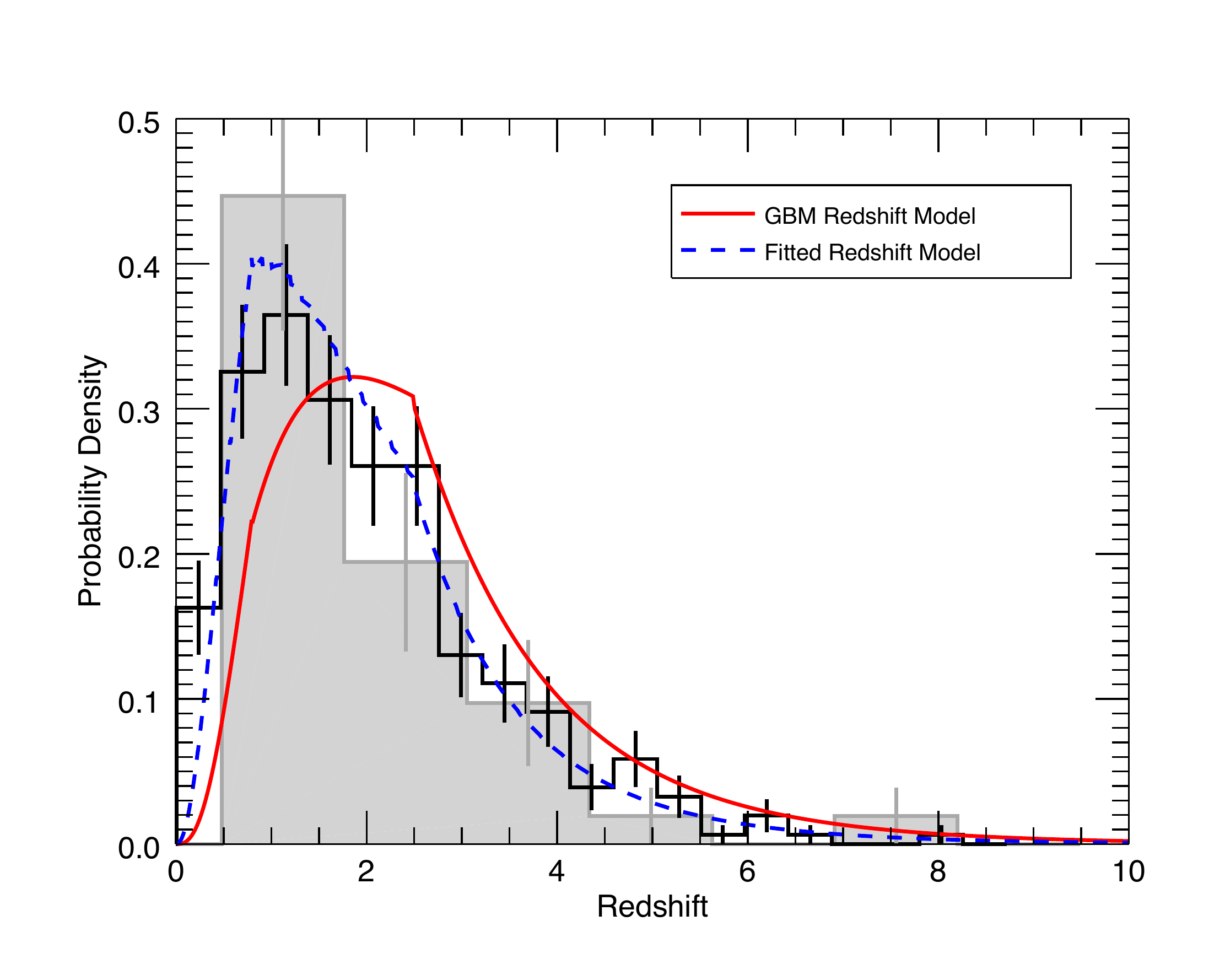}		
	\end{center}
\caption{The distribution of spectroscopic redshift for 335 long GRBs through March 2015 (black), and the distribution of 
redshifts of 40 long GBM GRBs (shaded gray).  The estimated redshift distribution of GRBs that can be triggered by GBM is 
shown in red.  The blue dashed line is the same redshift distribution model fit to the observed distribution of GBM GRBs with 
redshift.
\label{zDistrib}}
\end{figure}

Next, we test the method by comparing the same sample of GRBs with measured jet breaks to estimates of the $\theta_j$ 
derived assuming that we have not observed the redshift.  In this case, we sample from the GRB redshift distribution described 
by Equation~\ref{eq:zDistrib} and calculate the PDF for $\theta_j$ for each GRB.  Figure~\ref{thetaNoZ} shows that our 
estimation of $\theta_j$ is again largely consistent with the jet break estimates of $\theta_j$.  In the case of unknown redshift, 
the uncertainty on $\theta_j$ should be larger, and this is reflected in the comparison.  Also shown in Figure~\ref{thetaZ_noZ} 
is the estimated $\theta_j$ from the sample of 40 GBM GRBs with observed redshift compared to the estimation of $\theta_j$ 
assuming the redshift for those GRBs are unknown.  The comparisons are consistent, with a larger uncertainty in $\theta_j$ 
when the redshift is assumed unknown, as expected.  The centroids of the low-$z$ GRBs all lie below the line 
of unity due to the fact that they all have redshifts below the peak of the redshift distribution in Figure~\ref{zDistrib}.  Shown 
in Figure~\ref{theta_Z} is the dependence of $\theta_j$ on redshift for a given observation of $E_{\rm peak}$ and fluence, 
therefore a lower (higher) actual redshift for a given set of parameters would cause the derived $\theta_j$ to be higher (lower) 
than the estimate without knowledge of the redshift.  One should note that the 1$\sigma$ errors of the $\theta_j$ estimates 
describe the $\theta_j$ PDFs, which contain the full uncertainty from the redshift distribution.

\begin{figure}
	\begin{center}
		\subfigure[]{\label{thetaDistrib}\includegraphics[scale=0.40]{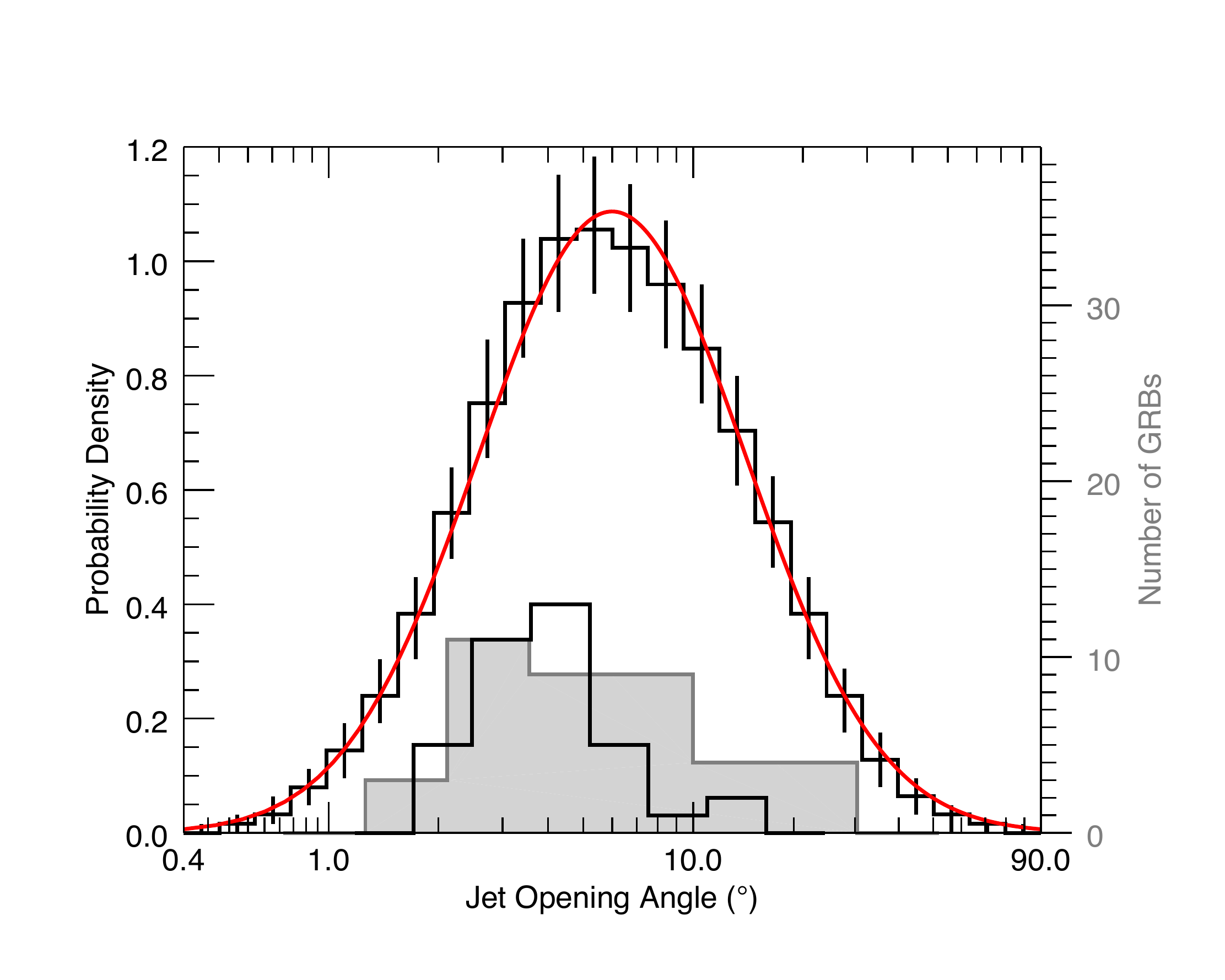}}
		\subfigure[]{\label{thetaPDF}\includegraphics[scale=0.40]{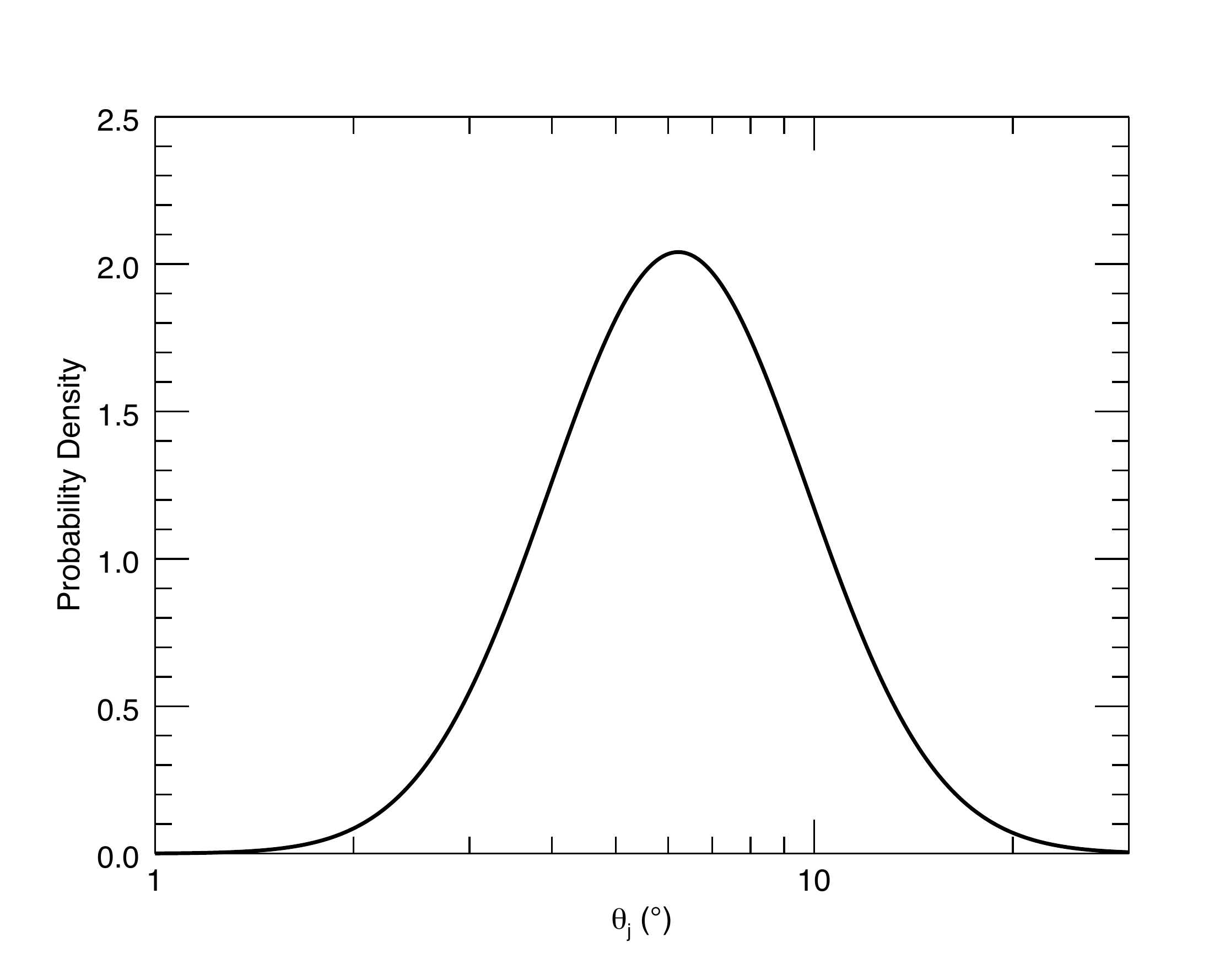}}\\
	\end{center}
	\caption{Panel~\ref{thetaDistrib} shows the distribution of derived jet opening angles for 638 GBM GRBs from the first 4 
	years of operation that are well fit by a Band function or a Comptonized function.  The distribution peaks at $\sim$6 
	degrees.  The small black histogram represents the distribution of inferred angles from observed jet breaks, and the 
	shaded histogram represents the distribution of derived angles from GBM GRBs with known redshift. 
	Panel~\ref{thetaPDF} shows a typical example of $\theta_j$ for a single GRB.  In this example the 68\% credible interval 
	for $\theta_j$ is $4-10^\circ$.
	\label{thetaDistribution}}
\end{figure}

Similarly, the $\theta_j$ PDFs can be calculated for each of the GRBs in our GBM sample.  The PDFs are closely modeled 
as log-normal distributions and can be readily collected to form the largest sample to date to estimate the distribution of jet 
opening angles.  Figure~\ref{thetaDistrib} shows the histogram of $\theta_j$ for the GBM sample and the log-normal fit to the 
distribution.  The distribution of $\theta_j$ inferred from measured jet breaks is shown for comparison.  As has been previously 
been speculated, most long GRBs have highly collimated jets with opening angles $< 10^\circ$~\citep{Frail01, Bloom03, 
Guetta05}, and our distribution shows that most GRBs indeed have $\theta_j < 10^\circ$.  In fact, by our estimation, 90\% of 
long GRB jet angles are $< 20^\circ$.  On the other end of the distribution, we estimate that $\sim2\%$ of opening angles are 
$< 1^\circ$.  Figure~\ref{thetaPDF}  shows an average example of the PDF for an individual $\theta_j$ estimate.  Our 
distribution for $\theta_j$ is consistent with that found by~\citet{Ghirlanda05a}, where the distribution was also found to peak at 
$<10^{\circ}$ and very few $>50^{\circ}$.

\begin{figure}
	\begin{center}
		\subfigure[]{\label{fluenceTheta}\includegraphics[scale=0.70]{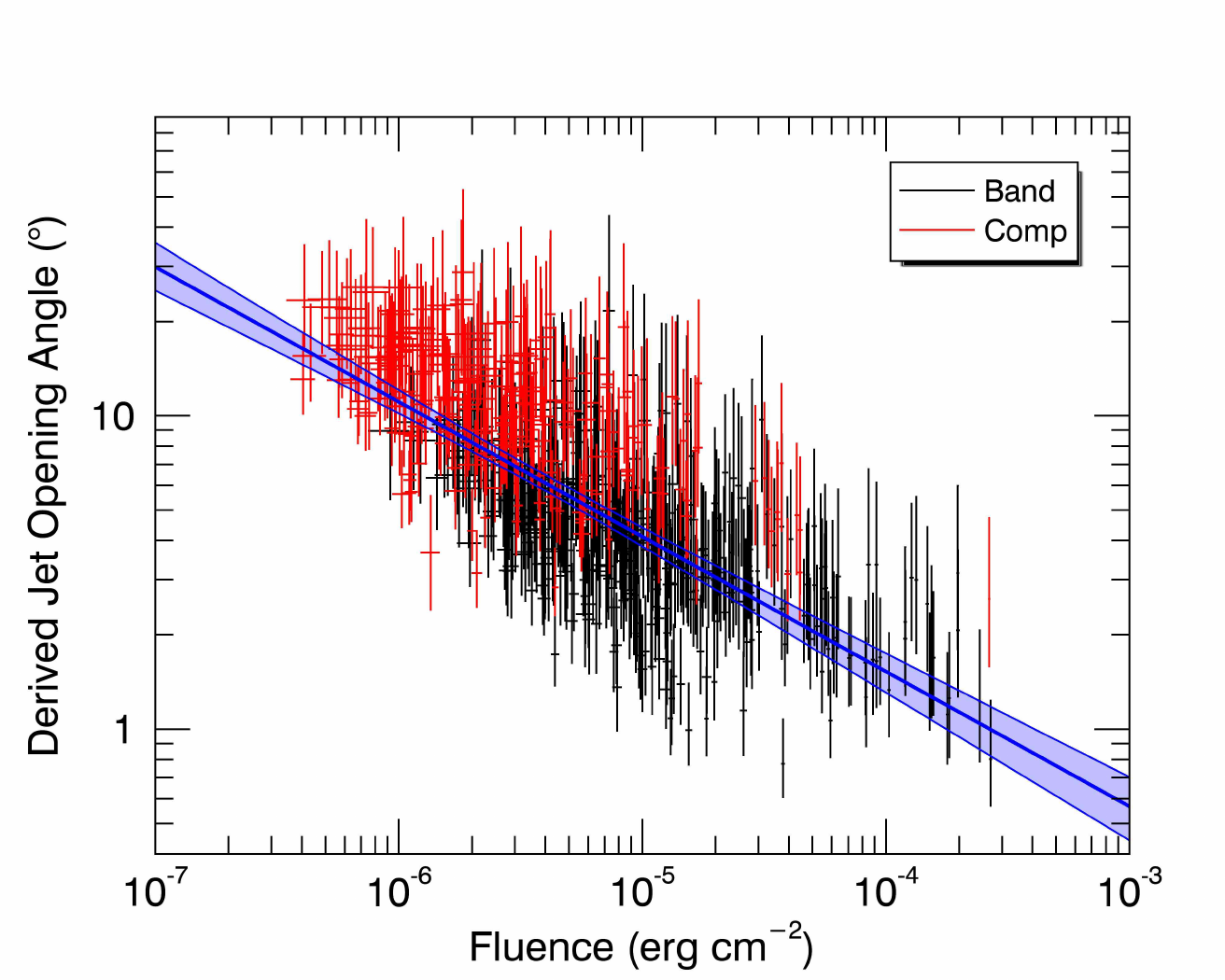}}\\
		\subfigure[]{\label{epeakTheta}\includegraphics[scale=0.70]{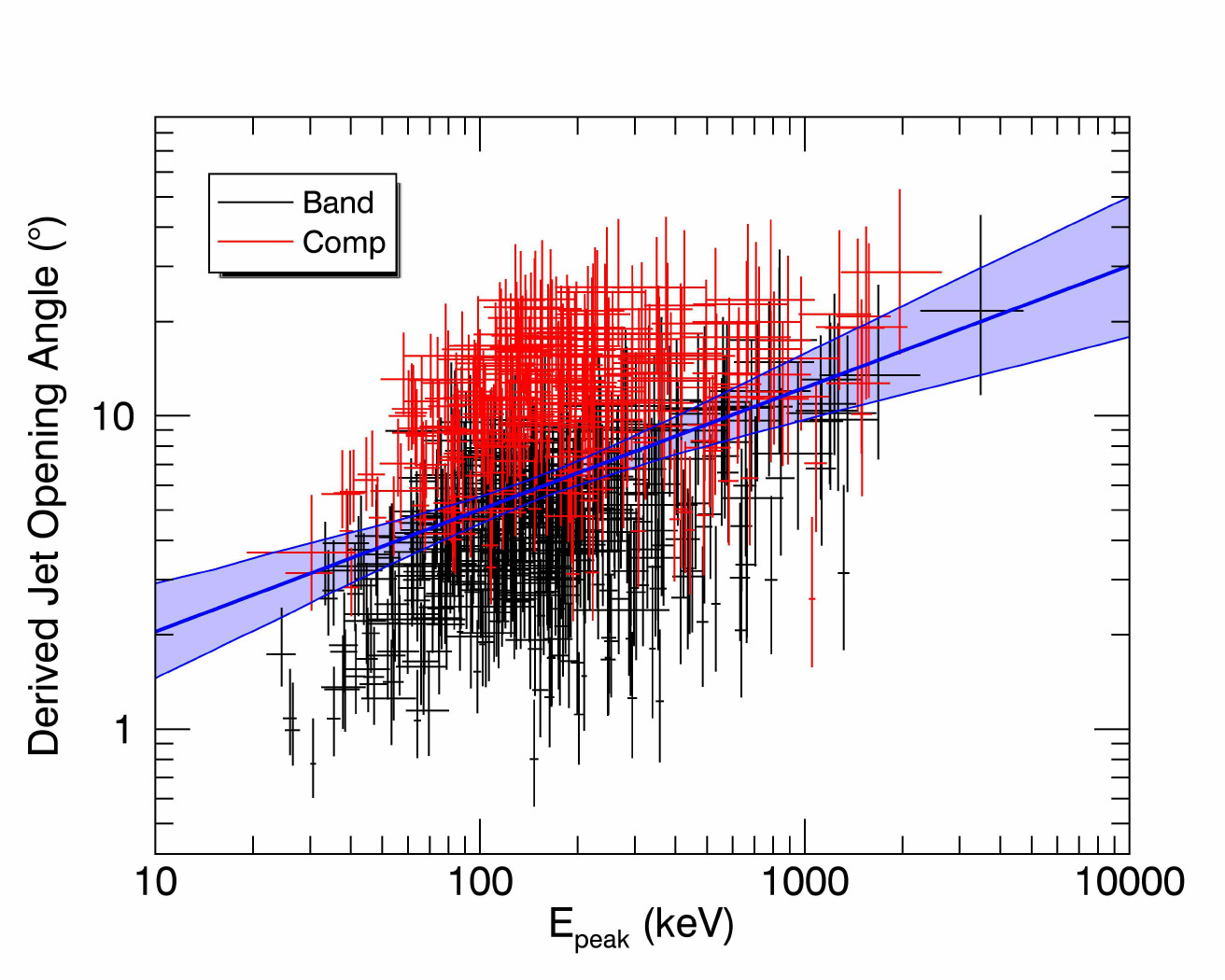}}
	\end{center}
	\caption{Panel~\ref{fluenceTheta} shows the dependence of the jet opening angle on GRB fluence.  The blue region is 
	the best fit power law regression, and if it is assumed to extend to lower fluence, a $90^\circ$ opening angle would be be 
	likely at $\sim (5-10)\times10^{-9} \ \rm erg \ cm^{-2}$.  Panel~\ref{epeakTheta} shows the dependence of the jet opening 
	angle on the measured GRB $E_{\rm peak}$.  If the regression is assumed to extend to higher values of $E_{\rm peak}$, 
	a $90^\circ$ opening angle would be likely at $\sim$~60--700~MeV.
	\label{thetaSystematics}}
\end{figure}

There are particular selection effects that can lead to truncation of our estimated distribution of $\theta_j$, primarily the 
limiting flux and fluence sensitivity of the detector and the potential existence of $E_{\rm peak}$ outside the GBM 
bandpass.  For example, we do not attempt to estimate $\theta_j$ for GRBs that have a poorly constrained $E_{\rm peak}$ or 
are generally too weak to be fit by a Band or Comptonized function.  Equation~\ref{eq:predictTheta} shows that $\theta_j$ will 
generally increase with decreasing fluence and increase with increasing $E_{\rm peak}$. In Figure~\ref{thetaSystematics} we 
show the correlations between our estimates of $\theta_j$ for the GBM sample and the fluence and $E_{\rm peak}$ for 
each GRB.  The lowest fluence in 10-1000 keV for the GRBs fit with a Band function is $9\times10^{-7}\ \rm erg \ cm^{-2}$, 
while the least fluent from the Comptonized function is $4\times10^{-7}\ \rm erg \ cm^{-2}$.  The GRBs in the 
GBM catalog that can only be well-fit by a power law range in fluence from $3\times10^{-7}-4\times10^{-5} \ \rm erg \ cm^{-2}$,  
therefore it is unlikely that the low fluence from the simple power-law fits alone would cause the GRBs that we disregarded to 
have significantly different jet angles. It is also unlikely because a regression indicates that $\theta_j$ would approach $90^
\circ$ at one to two orders of magnitude lower fluence than what has been observed with GBM.  Alternatively, Figure~
\ref{epeakTheta} shows that an $E_{\rm peak}$ that has been redshifted below the GBM bandpass could indicate that $\theta_j 
< 1^\circ$.  GBM has observed a handful of GRBs with $E_{\rm peak} > 1$ MeV, but none have been observed to  approach 
the 40 MeV upper bound of the detector band.  A regression indicates that an $E_{\rm peak}$ above 40 MeV would likely 
approach a jet angle of $90^\circ$.

In addition to observational selection effects, changes to our assumed GRB redshift distribution can affect the estimated $
\theta_j$ distribution.  We have found that minor changes to the GRB redshift model do not significantly affect our $\theta_j$ 
and energetics distributions.  For example, the difference between the redshift distribution used in this paper and the redshift 
distribution derived for $\Swift$in~\citet{Coward13} produces a difference of $\sim0.1^\circ$ in the peak of the $\theta_j$ 
distribution and no change in the width.  Additionally, we produced $\theta_j$ estimates from the redshift model that we 
fitted to the observed GBM redshift distribution (blue dashed line in Figure~\ref{zDistrib}).  The $\theta_j$ estimates using this 
redshift distribution changed by an average of 12\% or 0.4$\sigma$ compared to our preferred redshift distribution.  This 
causes a $0.5^\circ$ shift in the peak of the ensemble $\theta_j$ distribution in Figure~\ref{thetaDistrib}.  These comparisons 
show that this method of estimating $\theta_j$ is robust and is insensitive to moderate changes in the assumed redshift 
distribution.

\subsection{Rest-Frame Energetics}
\begin{figure}
	\begin{center}
		\subfigure[]{\label{eiso}\includegraphics[scale=0.40]{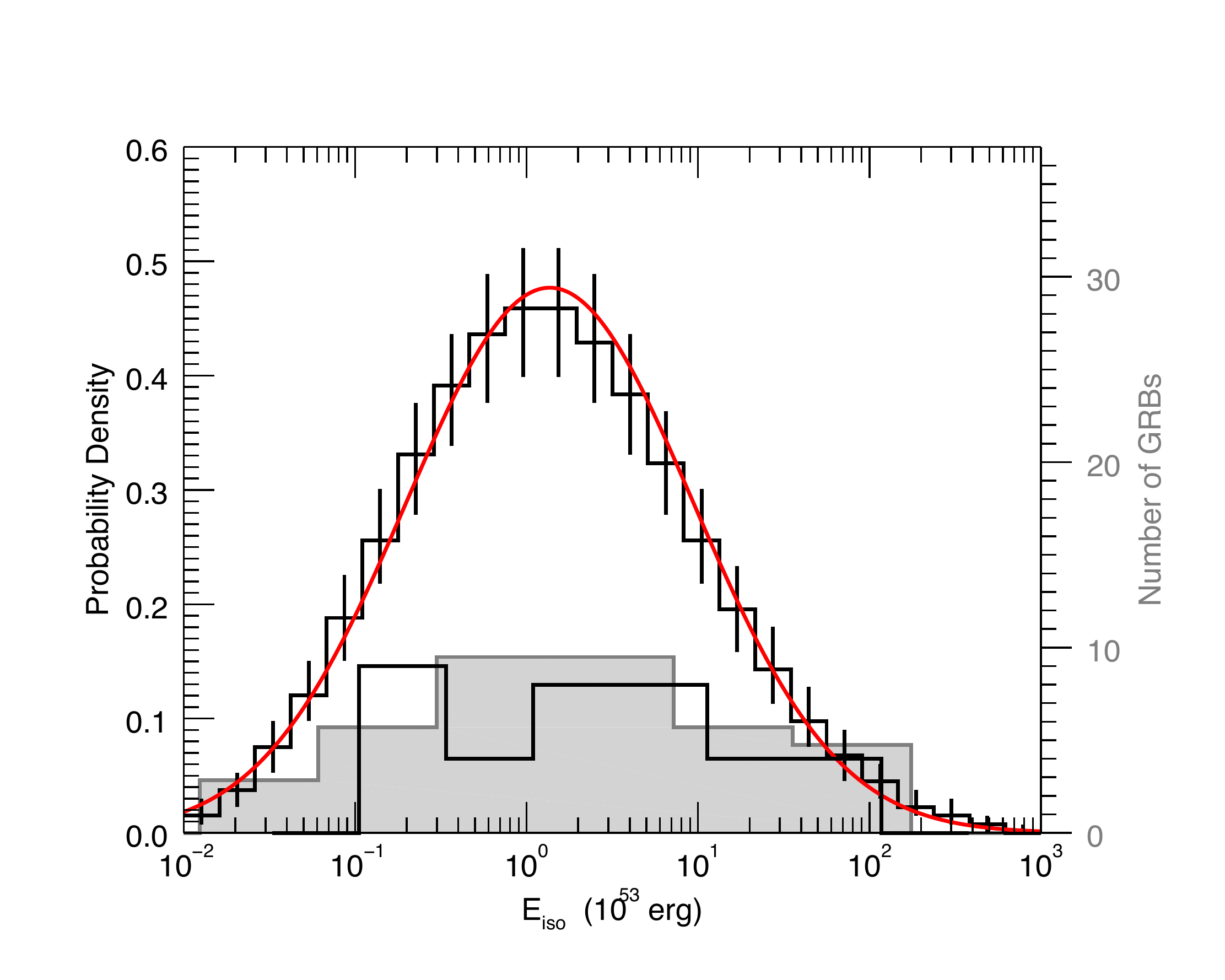}}
		\subfigure[]{\label{egamma}\includegraphics[scale=0.40]{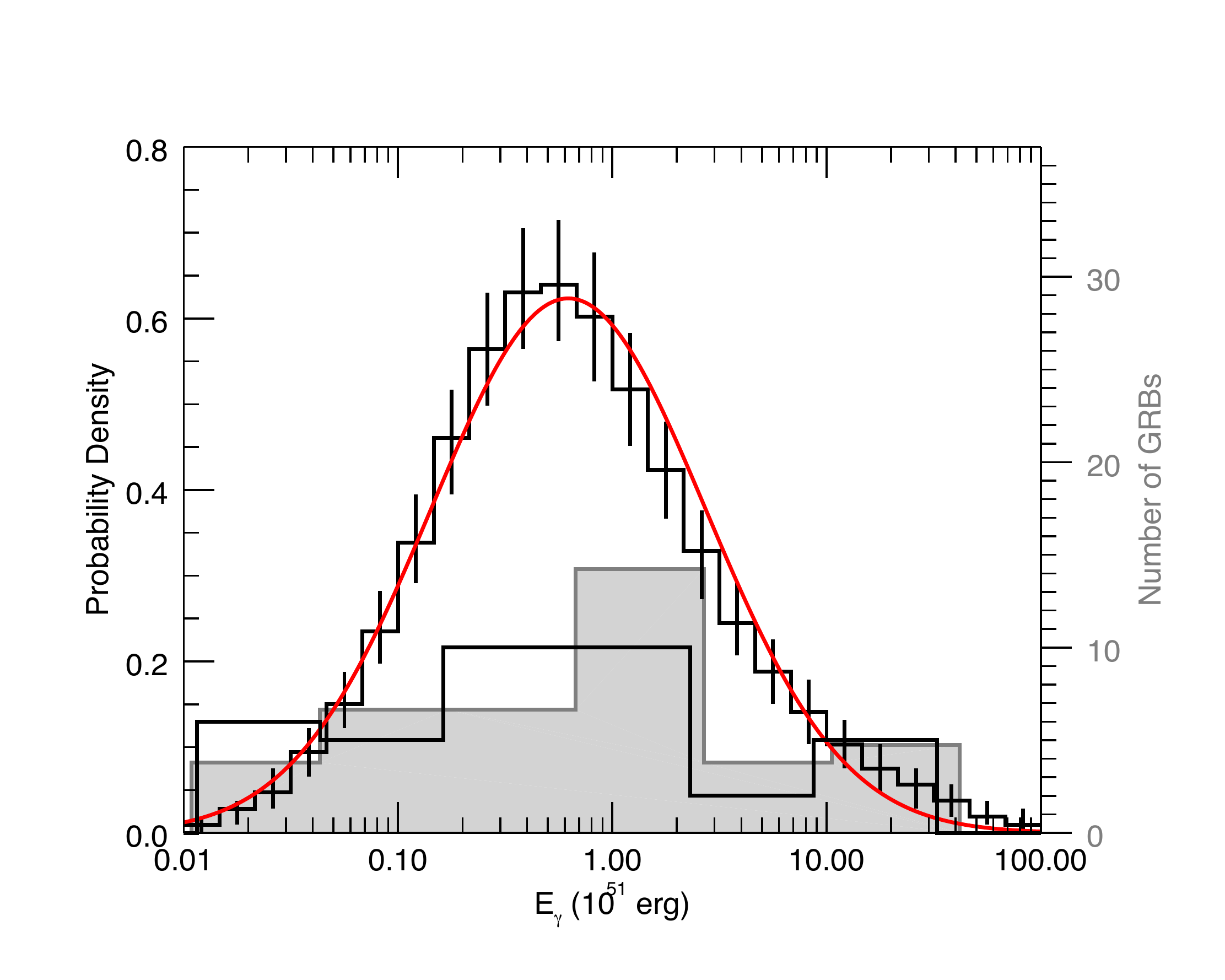}}
	\end{center}
	\caption{Panel~\ref{eiso} shows the distribution of the isotropic rest-frame energy in gamma rays, which peaks at 
	$\sim1\times10^{53}$ erg.  Panel~\ref{egamma} shows the distribution of the collimation-corrected rest-frame energy in 
	gamma rays.  The distribution peaks at $\sim6\times10^{50}$ erg.  The small black histograms represent the distribution 
	of measured $E_{\rm iso}$ and $E_\gamma$ from the inferred jet opening angles, and the shaded histograms represent 
	the distribution of derived $E_{\rm iso}$ and $E_\gamma$ from GBM GRBs with known redshift.
	\label{energyDistributions}}
\end{figure}

Similar to the estimation of the $\theta_j$ PDFs for GRBs by sampling from the GBM-observable redshift distribution, we can 
estimate the $E_{\rm iso}$ PDFs (1 keV--10 MeV) for our sample of GBM GRBs.  Additionally, we can use our estimates of 
$\theta_j$ for each GRB to produce PDFs for the collimation-corrected energy, $E_\gamma$.  In this case, we sample from the 
joint redshift--$\theta_j$ distribution to accurately calculate the $E_\gamma$ PDF.  In practice, we first sample from the redshift 
distribution, and then we sample from the $\theta_j$ distribution conditional on the sampled redshift.  Using this process, we 
estimate the $E_{\rm iso}$ and $E_\gamma$ PDFs for our GBM sample and construct the histograms, shown in 
Figure~\ref{energyDistributions}.  Our results show that the $E_{\rm iso}$ distribution is broader than the $E_\gamma$ 
distribution, although we show that $E_\gamma$ likely spans 4 orders of magnitude and appears to have an interesting 
non-Gaussian high-energy tail.  The distribution of $E_{\rm iso}$ appears to peak at $\sim1\times10^{53}$ erg and has 
relatively few events at $> 1\times 10^{55}$ erg, which places the most energetic observed GRB to date, 080916C~
\citep{Abdo09}, at the 98th percentile of all likely events.  Using the estimated jet break from~\citet{Maselli14} for the brightest 
observed GRB 130427A, $E_\gamma \approx 7\times10^{50}$ erg, which is only at the 46th percentile of our distribution.  
However, the famous `naked eye' burst 080319~\citep{Racusin08} is estimated to currently have the largest 
collimation-corrected bolometric gamma-ray energy (1 kev--10 MeV) at $\approx 1\times10^{52}$ erg, based on the 
Konus-Wind gamma-ray data~\citep{Golenetskii08}, which is at $\sim$98th percentile for the $E_\gamma$ distribution.  Our distribution of $E_{\rm iso}$ is broadly consistent with previous observations~\citep{Frail01, Ghirlanda04, Amati06, 
Nava12}.  Our $E_\gamma$ distributions are also similar to the observations from~\citet{Frail01} and~\citet{Ghirlanda04}.
 
\begin{figure}
	\begin{center}
		\subfigure[]{\label{liso}\includegraphics[scale=0.40]{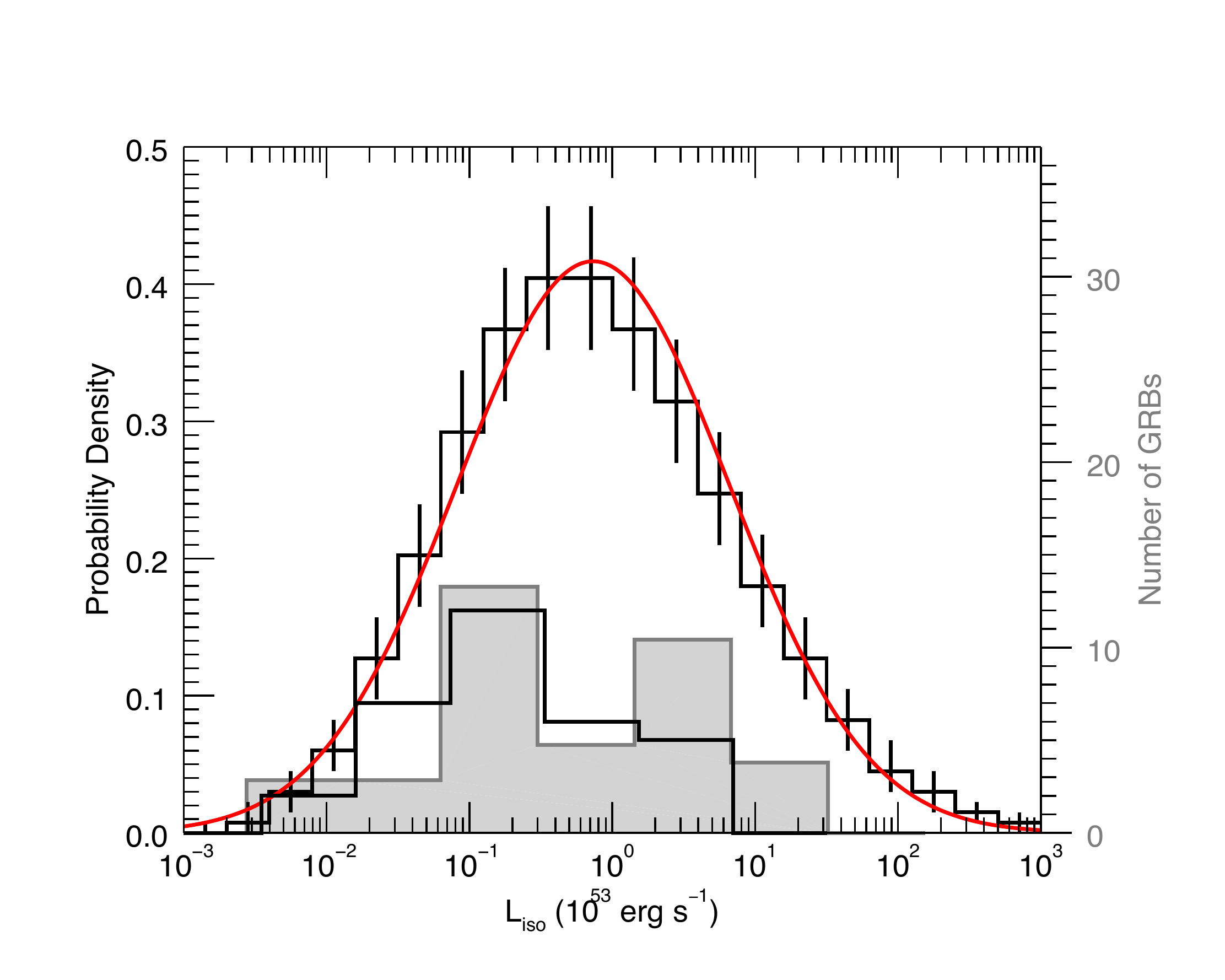}}
		\subfigure[]{\label{lgamma}\includegraphics[scale=0.40]{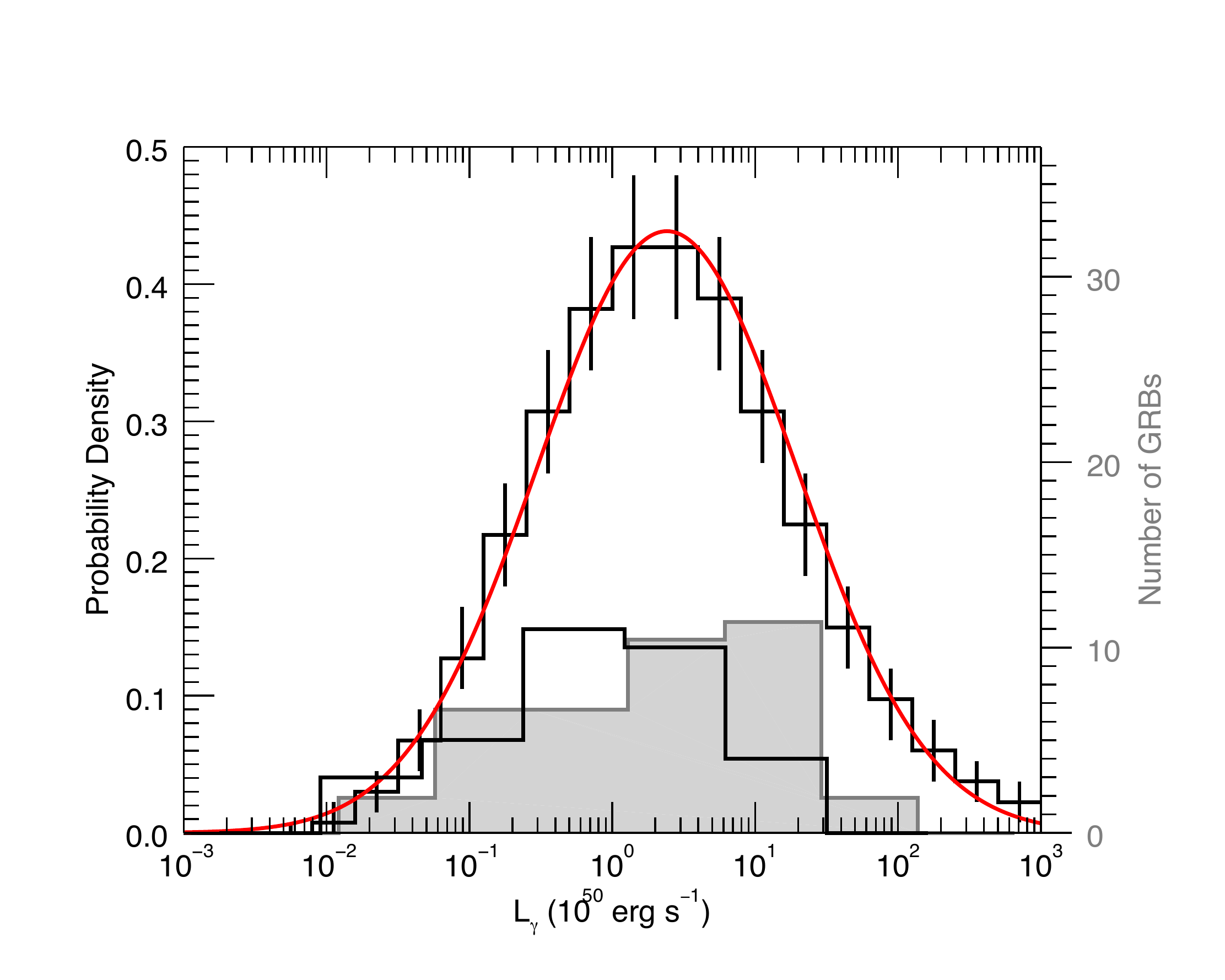}}
	\end{center}
	\caption{Panel~\ref{liso} shows the distribution of the isotropic rest-frame peak luminosity (derived from the 1 s observed 
	peak flux) in gamma rays, which peaks at $\sim8\times10^{52} \ \rm erg \ s^{-1}$.  Panel~\ref{lgamma} shows the 
	distribution of the collimation-corrected rest-frame peak luminosity in gamma rays.  The distribution peaks at 
	$\sim2\times10^{50}$ erg $\rm s^{-1}$.  The small black histograms represent the distribution of measured $L_{\rm iso}$ 
	and $L_\gamma$ from the inferred jet opening angles, and the shaded histograms represent the distribution of derived 
	$L_{\rm iso}$ and $L_\gamma$ from GBM GRBs with known redshift.\label{luminosityDistribution}}
\end{figure}

We perform the same calculations to estimate the peak luminosity distributions, $L_{\rm iso}$ and $L_\gamma$, as we did for 
the rest-frame energy.  In these calculations, instead of using the time-integrated spectrum for each GRB, we use the spectrum 
at the 1 s peak of each GRB.  Because the spectrum at the peak of the GRB is not always as well constrained as the 
 time-integrated spectrum, there are a number of GRBs in our GBM sample that do not have either an acceptable Band or 
Comptonized fit, and so we do not include those GRBs in the luminosity estimations.  In total there are 445 GRBs (311 Band 
and 134 Comptonized) from the GBM sample that have estimated peak luminosities.  Our estimated distributions of $L_{\rm 
iso}$ and $L_\gamma$ are shown in Figure~\ref{luminosityDistribution}.  Similar to what we find with the rest-frame energy, 
$L_\gamma$ has a narrower distribution than $L_{\rm iso}$, which is due to the fact that there exists a distribution of $\theta_j
$.  Our distributions suggest that while isotropic peak luminosities may approach and exceed $10^{55}$ erg $\rm s^{-1}$ in 
some cases, the actual rest-frame peak luminosity when corrected for collimation rarely exceeds $10^{52}$ erg $\rm s^{-1}$.  
The $L_{\rm iso}$ distribution presented here is in good agreement with the distribution presented in~\citet{Nava12}.  We 
estimate that 130427A, although likely the brightest GRB observed to date, is only in the top third in isotropic luminosity 
and near the median in $L_\gamma$.

\begin{figure}
	\begin{center}
		\subfigure[]{\label{fluenceEpz}\includegraphics[scale=0.45]{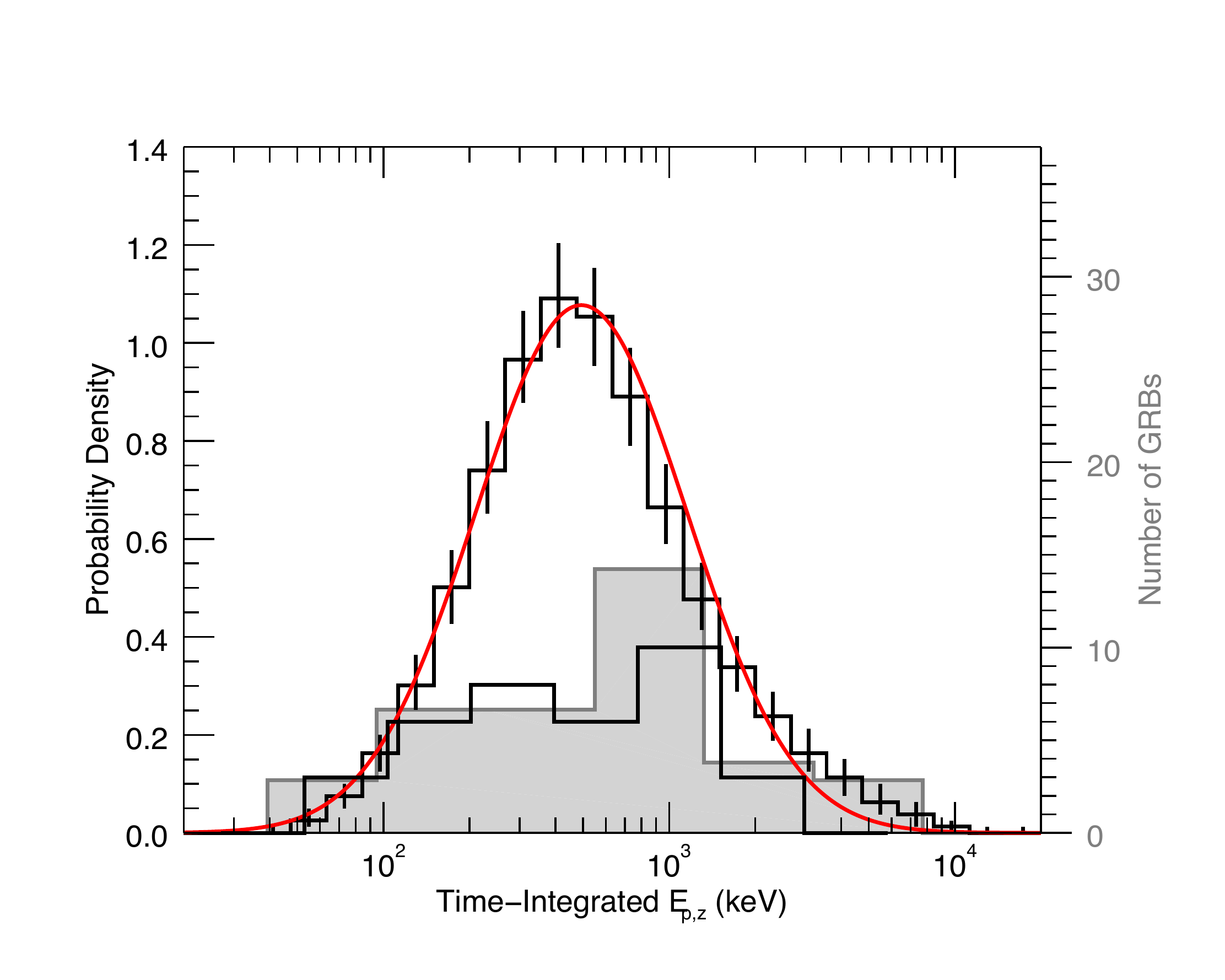}}
		\subfigure[]{\label{peakEpz}\includegraphics[scale=0.45]{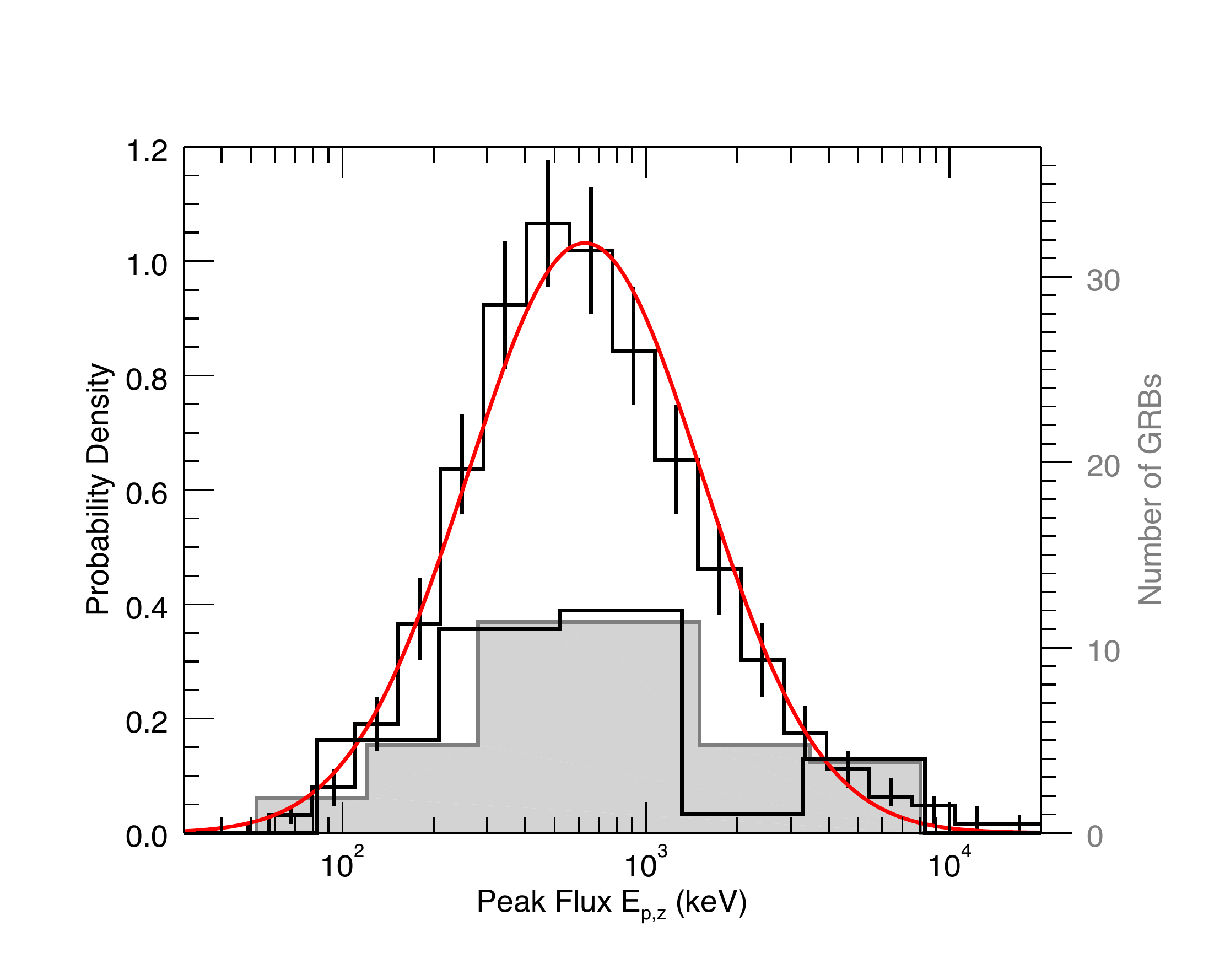}}
	\end{center}
	\caption{Distribution of estimated rest-frame $E_{peak}$. The time-integrated $E_{p,z}$ peaks at $\sim$500 keV, and the 
	peak flux $E_{p,z}$ peaks at $\sim$600 keV.  The small black histograms represent the distributions of observed 	
	$E_{p,z}$ from the jet break sample, and the shaded histograms represent the distribution of derived $E_{p,z}$ from GBM 
	GRBs with known redshift.
	\label{EpzDistribution}}
\end{figure}

Many studies have looked at estimating the $E_{\rm peak}$ distribution in the rest-frame, investigating if there is a particular 
energy at which the intrinsic spectrum peaks~\citep{Mallozzi95, Schaefer03, Liang04, Collazzi11}.  We can add to this 
investigation by estimating the rest-frame distribution of $E_{\rm peak}$.  Figure~\ref{EpzDistribution} shows the distributions of 
the time-integrated $E_{p,z}$ and $E_{p,z}$ at the peak of the GRB.  The distributions generally peak between 500--600 keV 
and have slightly non-Gaussian high-energy tails, which may hint at a truncation of the $E_{p,z}$ distribution, particularly at 
low energy.  The low-energy end of our distributions imply that the GBM bandpass would impose restrictions on the observed 
distributions for GRBs at $z\gtrsim9$.  The paucity of GRBs with $E_{p,z} > 10$ MeV indicates that the upper threshold of the 
GBM band does not impose a restriction on observed $E_{\rm peak}$, due to the fact that a higher energy $E_{p,z}$ would 
imply a larger energy and luminosity and would be even more likely to be observed by GBM than sub-MeV $E_{p,z}$.  The 
$E_{p,z}$ distributions are broadly interpreted as being defined by both the emission physics within the jet and the bulk 
Lorentz factor which blue-shifts the $E_{p,z}$ from the comoving jet frame to the cosmological rest-frame. The spread 
in the distributions may be attributed primarily to the differences in magnetic field strength and dynamics and to the distribution 
of bulk Lorentz factors among the GRBs~\citep{Baring04, Burgess14}.

Finally, to quantify the accuracy of the estimation of the rest-frame energetics employing the Ghirlanda relation and a 
proposed redshift distribution, we compare our estimates of the energetics to the sample of calibration GRBs in 
Tables~\ref{GRBLoZ}~and~\ref{GRBHiZ}.  Out of the 37 GRBs in our calibration sample, 59\% (97\%) are consistent within 1$
\sigma$ (2$\sigma$) for the estimation of $E_{iso}$,   65\% (95\%) are consistent within 1$\sigma$ (2$\sigma$) for the 
estimation of $E_\gamma$, and 68\% (100\%) are consistent within 1$\sigma$ (2$\sigma$) of the time-integrated $E_{p,z}$.  
Similar numbers are found in the comparison with the luminosity.  These comparisons indicate that our method provides 
accurate estimates that are well-calibrated to the calculation of the energetics for GRBs with known redshift.

\subsection{Correlations}
\begin{figure}
	\begin{center}
		\subfigure[]{\label{EpzEiso}\includegraphics[scale=0.60]{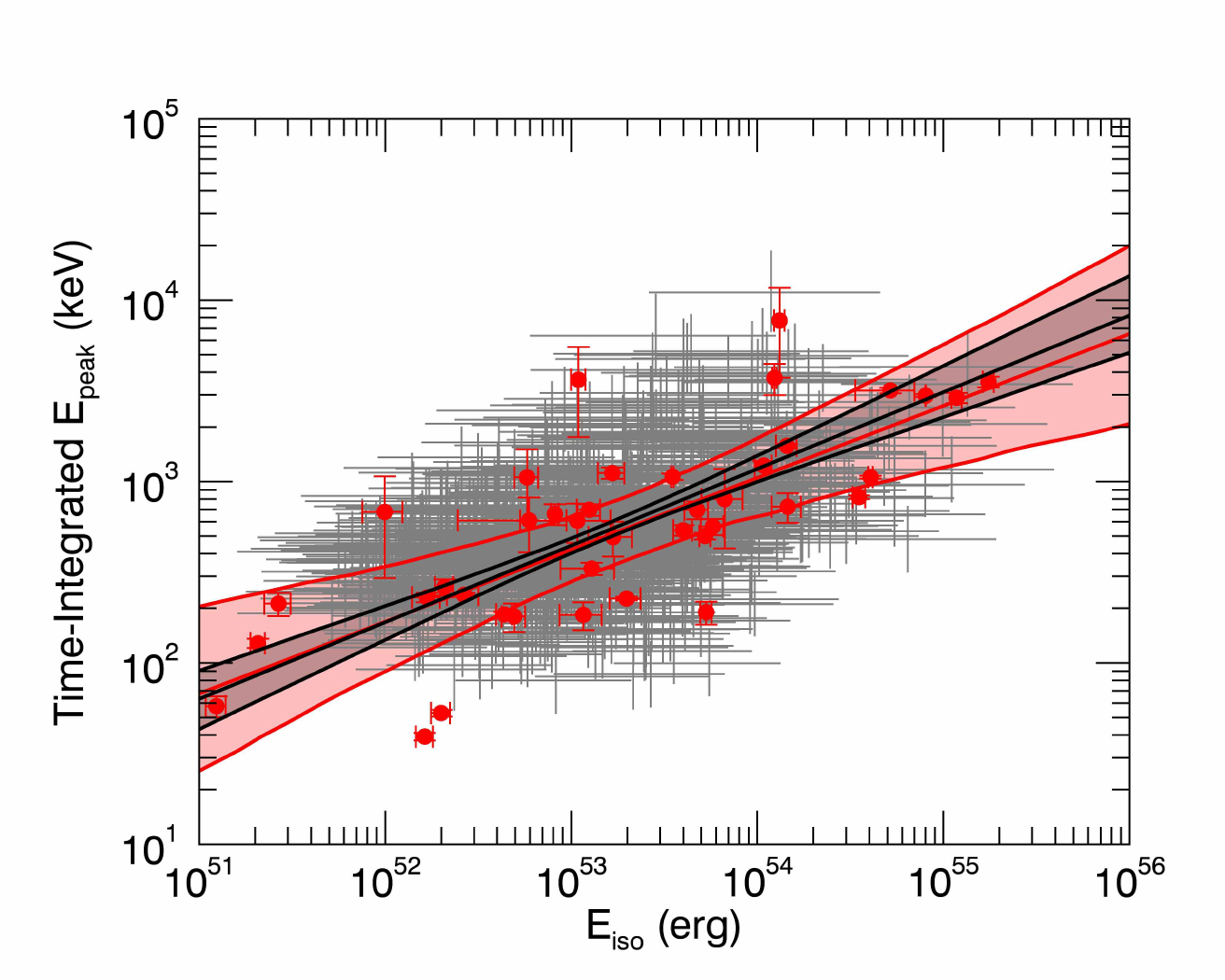}}
		\subfigure[]{\label{EpzLiso}\includegraphics[scale=0.60]{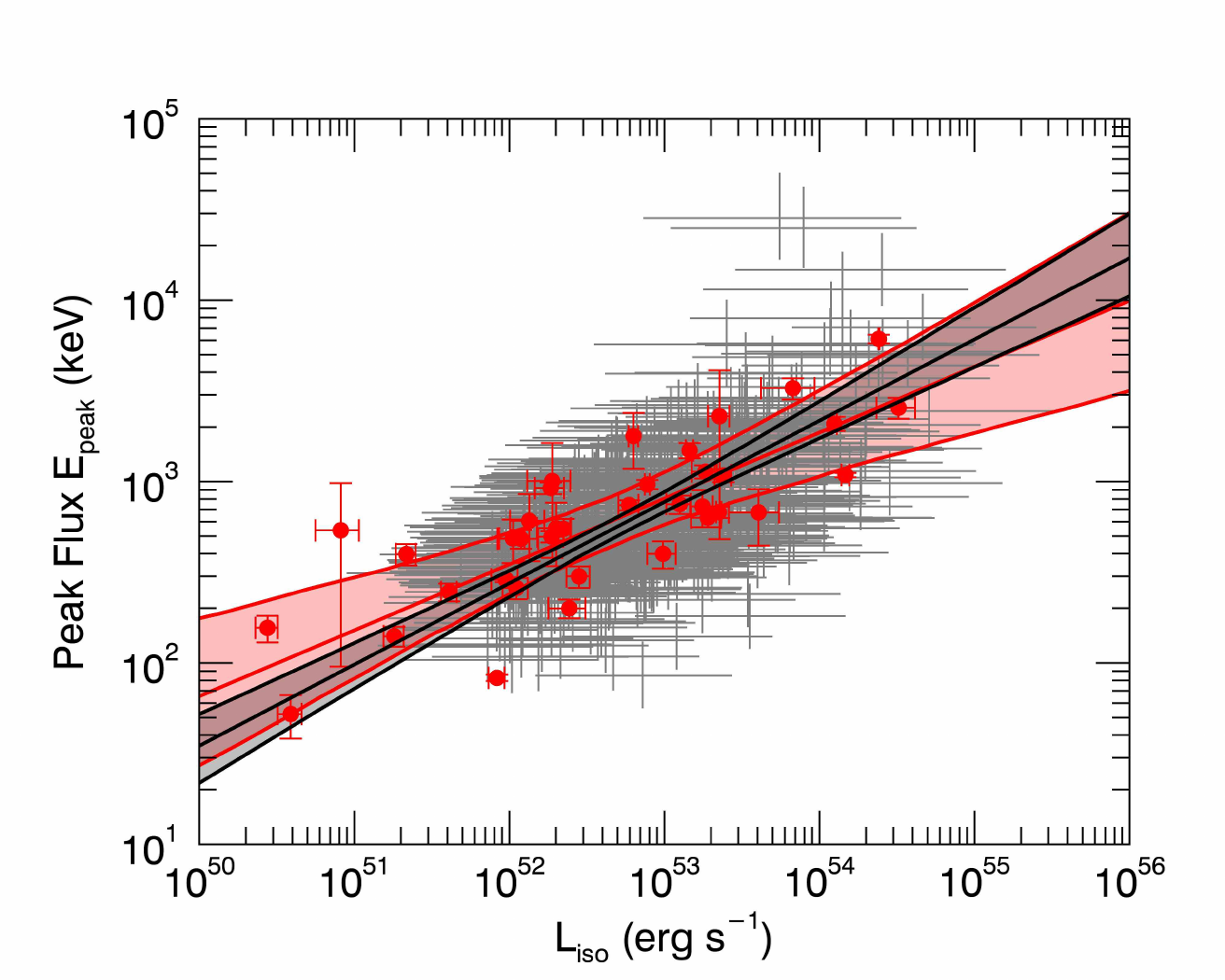}}\\
		\subfigure[]{\label{EpzLgamma}\includegraphics[scale=0.60]{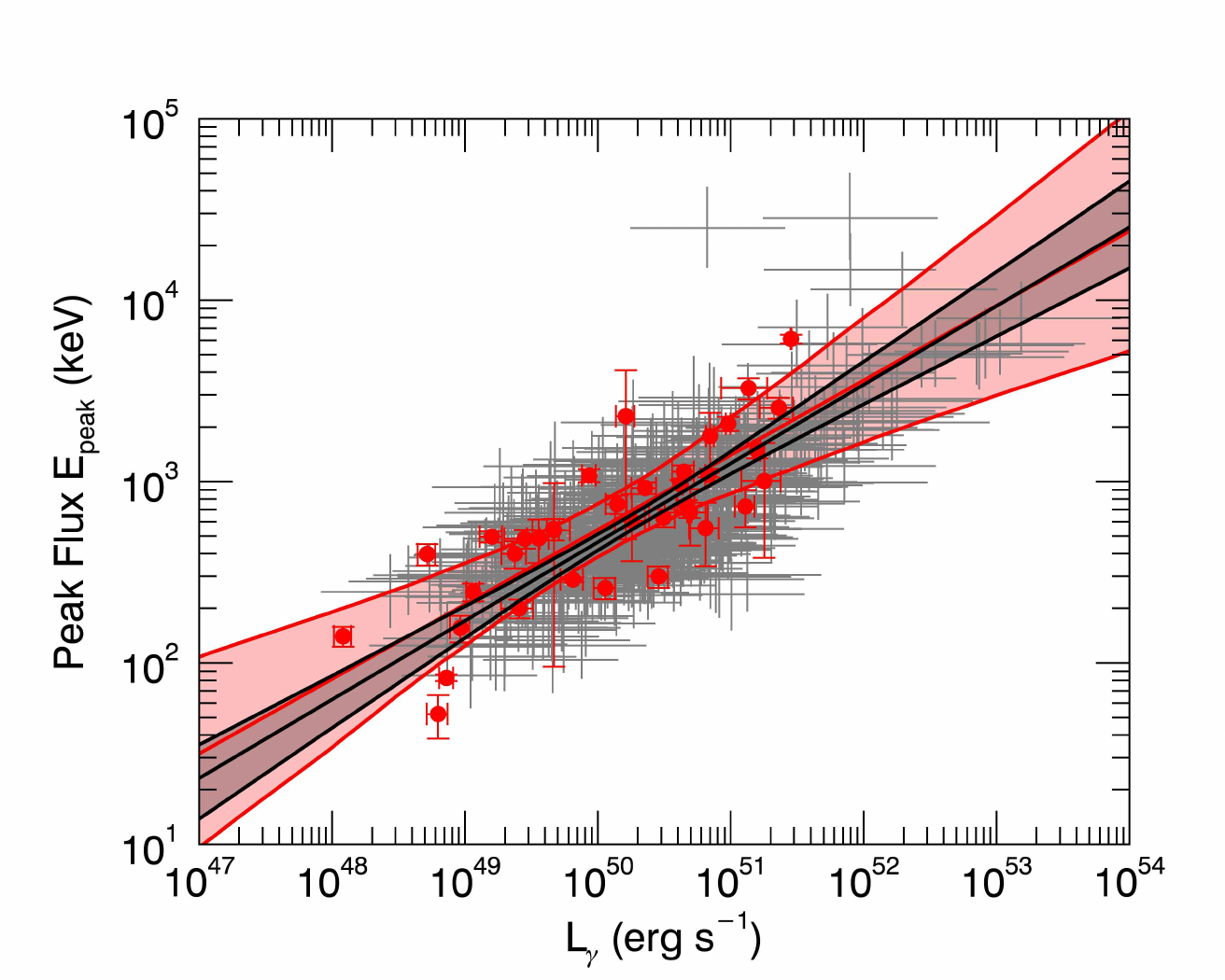}}
	\end{center}
	\caption{The (a) $E_{p,z}-E_{\rm iso}$, (b) $E_{p,z}-L_{\rm iso}$, and (c) $E_{p,z}-L_{\gamma}$ correlations using the 
	GBM sample.  The error bars denote the 68\% credible regions for each data point.  The gray regions are the 99\% 
	credible regions for the power-law fits to the large population of GRBs with unknown redshift (gray points),  and the red 
	regions are the 99\% credible regions for the power-law fits to the GBM GRBs with known redshift (red points).  
	\label{correlations}}
\end{figure}

Now that we have calculated the rest-frame energy, luminosity, and $E_{\rm peak}$, we investigate a few of the rest-frame 
correlations that have been discussed in literature.  We use the large number of GRBs from our GBM sample to plot the 
time-integrated $E_{p,z}-E_{\rm iso}$~\citep{Amati02}, the peak $E_{p,z}-L_{\rm iso}$~\citep{Yonetoku04}, and the peak 
$E_{p,z}$-$L_\gamma$~\citep{Ghirlanda05b} correlations, shown in Figure~\ref{correlations}.  We fit each correlation with a 
power law to find the best-fit normalization and power-law index.  Note that significant outliers to the correlations are likely to 
be GRBs that are at the extreme tails of the redshift distribution in Figure~\ref{zDistrib}.

For the Amati relation, using the large sample of GRBs without known redshift, we find the best fit correlation to be
\begin{equation}
	\frac{E_{p,z}}{1\ \rm keV} = 10^{2.71\pm0.01}\biggl( \frac{E_{\rm iso}}{1.41\times10^{53} \ \rm erg} \biggr)^{0.44\pm0.02}.
\end{equation}
We compare this to the best-fit power law for the GBM GRBs with known redshift, which gives an index of $0.40\pm0.05$.  The power law indices from both samples are consistent with each other and are roughly consistent (within $2-3\sigma$) with 
the theoretical prediction of~\citet{Levinson05} of 1/2 from simple annular jet dynamics and viewing angle effects.  It is 
apparent that the correlation has a large dispersion, even when accounting for uncertainties, which makes it difficult to use to 
study cosmology as has been previously discussed~\citep{Nakar05, Band05, Collazzi12}.

A more narrow correlation is the peak $E_{p,z}-L_{\rm iso}$ correlation, which we find is best fit by the power law
\begin{equation}
	\frac{E_{p,z}}{1\ \rm keV} = 10^{2.83\pm0.02}\biggl( \frac{L_{\rm iso}}{7.5\times10^{52} \ \rm erg \ s^{-1}} 
	\biggr)^{0.45\pm0.02}.
\end{equation}
We find the correlation slope is less steep than, but close to, $\sim0.5$ that was first fit by~\citet{Yonetoku04}.  The 
best-fit correlation using the GBM redshift GRBs gives an even shallower index of $0.36\pm0.1$ but is consistent with the 
larger distribution at 1$\sigma$.  It is also interesting to note that three of the GRBs with known redshift that exist at 
low-luminosity compared to the sample of unknown redshift have an associated supernova: GRBs 081007~\citep{Zhi-Ping13}, 
091127~\citep{Cobb10}, and 101219B~\citep{Ugarte11}.

We also find that the tightest of the three correlations is the peak $E_{p,z}-L_\gamma$ correlation, which is best described as 
\begin{equation}
	\frac{E_{p,z}}{1\ \rm keV} = 10^{2.83\pm0.01}\biggl( \frac{L_{\gamma}}{2.4\times10^{50} \ \rm erg \ s^{-1}} 
	\biggr)^{0.43\pm0.01}.
\end{equation}
\citet{Ghirlanda05b}, using a small sample of 16 GRBs with $E_{p,z}$ and $L_\gamma$, found a correlation slope of 0.56.  
We find, using the GBM redshift sample, that the index is $0.41\pm0.10$, which is also inconsistent with the findings 
of~\citet{Ghirlanda05b}.  The differences may originate from the small sample size in~\citet{Ghirlanda05b} and in the fact that 
they fit the correlation only considering the scatter in $L_\gamma$ instead of the scatter perpendicular to the power law fit.  It is 
also worth noting that three GRBs that have significant scatter from the correlation are the two SN-associated low-luminosity 
GRBs 081007 and 091127 and the high-luminosity GRB~090902B~\citep{Abdo09b}, which has an additional power law 
spectral component spanning from keV to GeV.

\subsection{Jet Break Time Distribution}
\begin{figure}
	\begin{center}
		\subfigure[]{\label{tjDistrib}\includegraphics[scale=0.45]{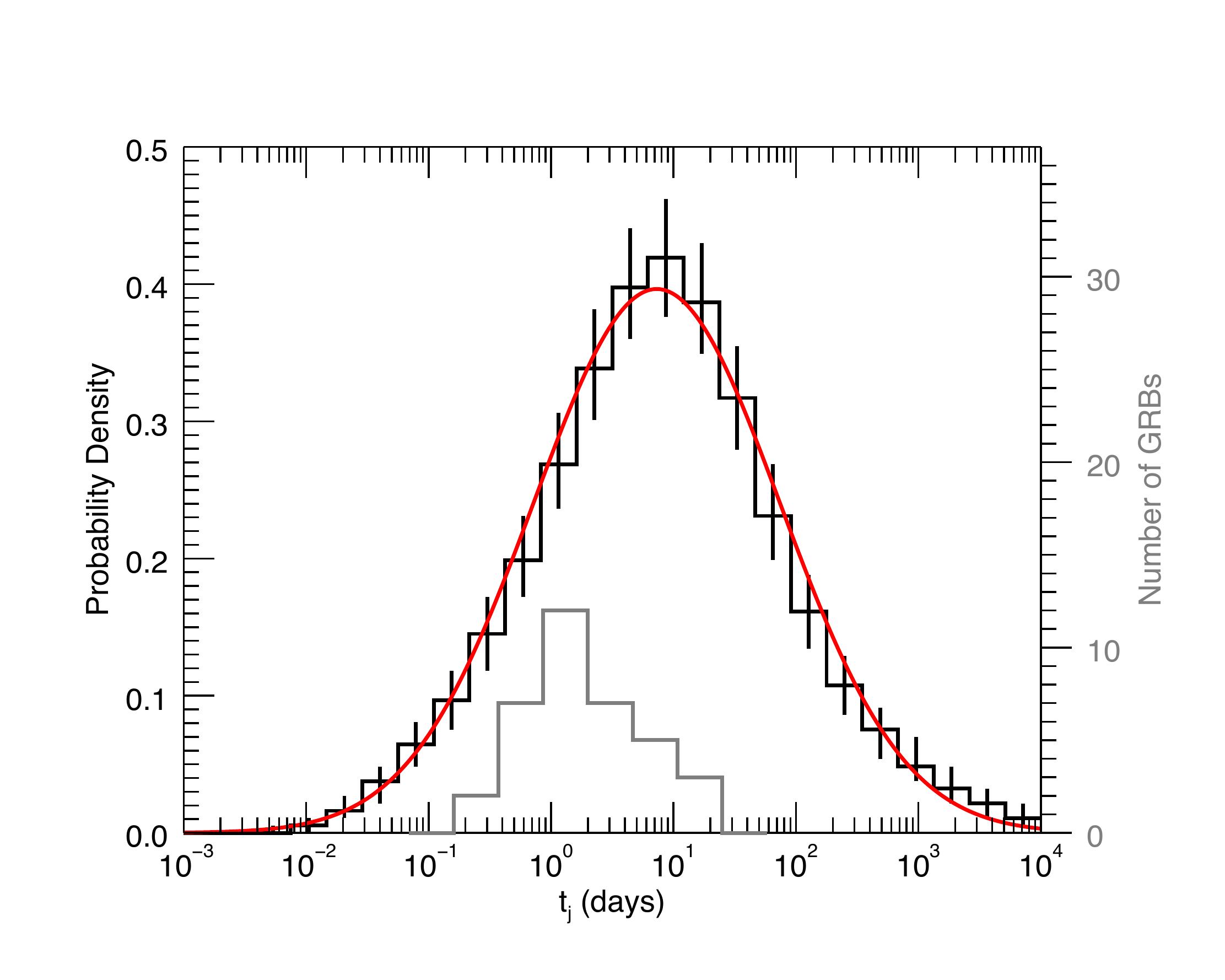}}
		\subfigure[]{\label{tjPDF}\includegraphics[scale=0.45]{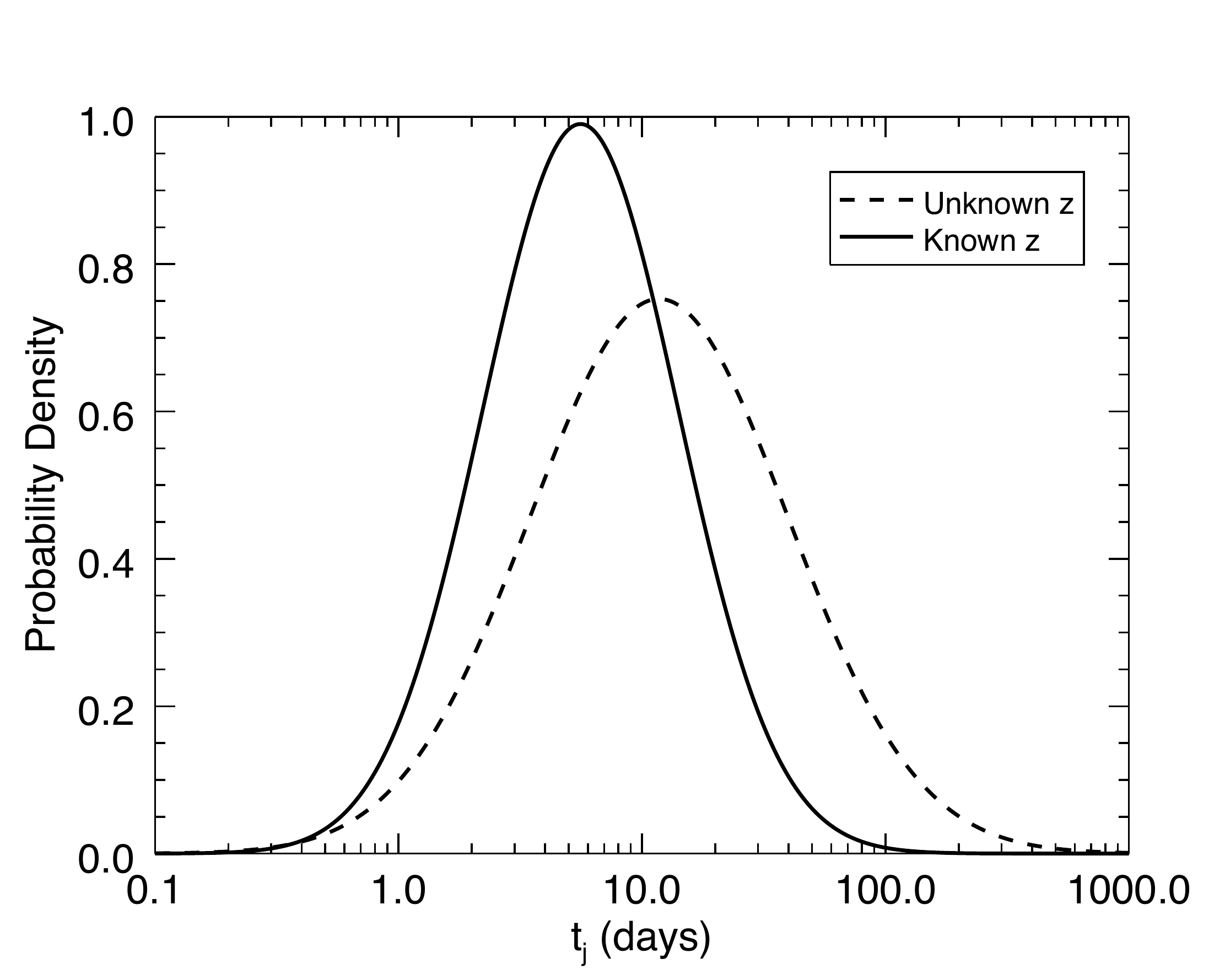}}
	\end{center}
	\caption{Panel~\ref{tjDistrib} shows the distribution of the estimated jet break time.  The distribution peaks at $\sim$7 
	days.  The gray histogram represents the distribution of observed jet break times from the afterglow. Panel~\ref{tjPDF} 
	shows a typical example of the $t_j$ PDF for a single GRB with known $z$ (68\% credible region of 6--14 days),
	and a typical $t_j$ PDF for a different GRB with unknown $z$ (68\% credible region of 4--42 days).\label{tjDistribution}}
\end{figure}

In addition to estimating the rest-frame collimation and energetics of a large sample of GRBs, we can use our estimates of $
\theta_j$ and $E_{\rm iso}$ to estimate the jet break time, $t_j$, for each GRB and uncover the distribution for $t_j$.  We invert 
Equation~\ref{eq:JetAngle} and calculate $t_j$ using the same assumed distributions for $\epsilon$ and $n_p$ as previously 
mentioned, and use Monte Carlo sampling of all PDFs in the equation.  The resulting distribution of $t_j$ is shown in 
Figure~\ref{tjDistribution}.  Most jet breaks have been observed from $\sim$0.5--10 days after the prompt emission.  Our 
distribution suggests that 90\% of observed jet break times can vary by $\sim$3 orders of magnitude, and a 
large fraction of jet breaks will not be observable.  If only {\it Fermi} observes the prompt emission of GRB, typically a detection 
by the {\it Fermi} LAT is required to trigger X-ray observations of the afterglow.  Since the LAT usually requires $\sim$12 hours 
to confirm detection and localize a GRB, this is the earliest that the afterglow for a {\it Fermi} GRB  would be observed.  Based 
on our $t_j$ distribution, we estimate that $\sim$10\% of GRBs have jet breaks that are within 12 hours after the prompt 
emission and are unlikely to be observed if only {\it Fermi} has observed the prompt emission.  If \Swift triggers on a GRB, the 
observed jet break distributions presented in~\citet{Racusin09} imply that the X-ray Telescope on \Swift can observe the jet 
break less than an hour after the prompt emission.  We estimate that only $\sim$1\% of jet breaks will occur less than an hour 
after the prompt emission, although in these cases it is important to have enough rapid afterglow before the jet break to 
adequately constrain the fit to the break.  The difficulties of observing rapid jet break affects only a small fraction of GRBs, but 
the situation at the high end of the $t_j$ distribution is more problematic.  We find that $\sim$44\% of GRBs will have $t_j > 10$ 
days and $\sim$13\% of GRBs will have $t_ j > 100$ days.  At these timescales, the afterglow flux will typically have faded 
below most X-ray and optical detector sensitivities and will be undetectable.  If an average $-1$ power-law 
decay in time is assumed for the X-ray and optical afterglow, then our $t_j$ distribution implies that an improvement of $
\sim1-2$ orders of magnitude in sensitivity is required to observe 85\% of all jet breaks.

Based on Figure~\ref{tjDistribution}, we estimate that at best only $\sim$50\% of GRB jet breaks will likely be 
detectable by current capabilities, and the fraction is certainly less when accounting for gamma-ray localization by GBM, timing 
of afterglow observations, and other afterglow observational constraints.  A previous study of \Swift afterglows~
\citep{Racusin09} found strong evidence of a jet break in the X-ray afterglow for only 12\% of GRBs and moderate evidence of 
a jet break for another 30\%.  A conclusion of that study was that at least 40\% of afterglows with missing jet breaks are due to 
observational biases which agrees with the results of our analysis.  Predicting the timing of the jet break without knowing the 
redshift by this method might not be practical either, since the $t_j$ PDF for an individual GRB is quite broad, as shown in 
Figure~\ref{tjPDF}.  Knowledge of the redshift will help narrow the PDF, although the 1$\sigma$ interval for predicting the jet 
break time will still usually be on the order of a few days.

\section{Summary}
In this paper we have described a method which can be used to estimate the jet opening angle of GRBs based on 
comparisons to estimates derived from observed jet breaks, even in cases where the redshift is not known. From the jet 
opening angle and redshift, the collimation-corrected energetics can be calculated.  We have shown that the cosmologically 
calibrated Ghirlanda relation is a tight correlation between the time-integrated $E_{p,z}$ and $E_\gamma$, and the correlation 
slope matches the estimate from the theoretical predictions.  By inverting the Ghirlanda relation, the jet opening angle can be 
estimated and is consistent with values inferred from observed jet breaks in the afterglow.  This estimate requires the fluence in 
gamma rays, the observed peak of the $\nu F_\nu$ spectrum, and the redshift of the GRB.  Furthermore, if the redshift is not 
known, we have shown that the modeled detector-dependent GRB redshift distribution can place constraints on $\theta_j$ and 
the rest-frame energetics of the GRB.  We also note that the $\theta_j$ and energetics PDFs and distributions produced for 
GRBs without known redshift are not sensitive to moderate changes in cosmological assumptions or the GRB luminosity 
function, therefore we do not expect our results to change significantly with a moderately different assumptions.

Combining the estimates of $\theta_j$ from the described method and the inferred rest-frame energetics of 638 long GRBs 
detected by {\it Fermi} GBM in its first 4 years of operation, we have produced distributions of rest-frame quantities that should 
provide insight into the progenitor and emission properties of collapsars.  We provide the parametrization of these 
distributions as well as our estimates of the energetics for all of the GRBs in our GBM sample.  These estimates 
represent the apparent radiative energetics from the jet of the GRB, and we still require estimation of the energy conversion 
efficiency and the Lorentz factor of the outflow to place constraints on the total energy budget of the GRBs. Additionally, three 
observed correlations between the rest-frame $E_{\rm peak}$ and the rest-frame energy were also shown.  We find that the 
$E_{p,z}-E_{\rm iso}$ correlation is roughly consistent with theoretical predictions based on simple jet dynamics and observing 
angle relative to the center of the jet.  We also find that our fits to the $E_{p,z}-L_{\rm iso}$ correlation is consistent with 
previous studies, however the slope for the $E_{p,z}-L_\gamma$ correlation is not consistent with the previous estimate of the 
slope by using GRBs with known redshift.  This inconsistency may be attributable to the relatively small sample size and 
different fitting method employed in the previous study.

Finally, we have estimated the distribution of jet break times for GRBs and have shown that a large fraction of jet breaks are 
currently not observable, which places considerable constraints on the ability to directly infer the jet opening angle via an 
observed jet break.  The jet break distribution has implications for follow-up observing strategies, future X-ray and optical 
detector sensitivities, and studying the many observational biases that may impact the non-detections of the jet breaks. 

\section{Acknowledgments}
A.G. is funded by the NASA Postdoctoral Program through Oak Ridge Associated Universities.

\clearpage


\clearpage

\begin{deluxetable}{ c  c  c  c  c  c  c  c  c }
\tablecolumns{9}
\tablewidth{0pt}
\tabletypesize{\scriptsize}
\setlength{\tabcolsep}{0.1in}
\tablecaption{GRB parameters  used to calculate jet opening angle ($z<1.5$)\label{GRBLoZ}}
\startdata
\toprule
\rule[-2pt]{0pt}{12pt}	\multirow{2}{*}{GRB} & \multirow{2}{*}{$z$} & $t_j$ & \multirow{2}{*}{$\alpha$} & \multirow{2}{*}{$\beta$} & 
$E_{\rm peak}$ & Fluence & Band & \multirow{2}{*}{$\rm Ref.^{\dagger}$} \\  & & (days) &  & & (keV) & $(10^{-6}\ \rm erg \ 
cm^{-2})$ & (keV) & \\
\midrule
970508 & 0.83 & $25.0\pm5.0$* & $-1.24\pm0.17$ & $-1.81\pm0.20$ & $432.6\pm261.2$ & $3.56\pm0.25$ & 20--2000 & 1,2\\
970828 & 0.96 & $2.20 \pm0.40$ & $-0.74\pm0.01$ &  $-2.07\pm0.40$ &  $298.0\pm30.0$ &  $96.0\pm 10.0$ &  20--2000 & 
1,3\\
980703 & 0.97 & $3.40\pm0.50$* & $-1.20\pm0.05$ & $-1.93\pm0.06$ & $280.1\pm31.7$ &  $39.8\pm0.90$ & 20--2000  & 1,2\\
990705 & 0.84 & $1.00\pm0.20$ &  $-1.05\pm0.21$ & $-2.20\pm0.10$ & $189.0\pm15.0$ & $75.0\pm8.0$ & 40--700 & 4\\
990712 & 0.43 & $1.60\pm0.30$ & $-1.88\pm0.07$ & $-2.48\pm0.56$ & $65.0\pm10.0$ & $6.50\pm0.30$ & 40--700 & 5,4\\
991216 & 1.02 & $1.20\pm0.40$ & $-1.20\pm0.01$ & $-2.22\pm0.02$ & $382.4\pm5.9$ & $174\pm0.53$ & 20--2000 & 1,2\\
010222 & 1.48 & $0.93\pm0.10$ & $-1.35\pm0.19$ & $-1.64\pm0.02$ & $309.0\pm12.0$ & $93.0\pm3.00$ &40--700 & 4\\
020405 & 0.69 & $1.67\pm0.52$ & $-1.10\pm0.40$ & $-1.87\pm0.23$ & $364.0\pm73.0$ & $74.0\pm7.00$ & 15--2000 & 5,4\\
020813 & 1.25 & $0.43\pm0.06$ & $-0.94\pm0.03$ & $-2.30\pm0.50$ & $142.0\pm14.0$ & $98.0\pm1.00$ & 2--400 & 4\\
030329 & 0.17 & $0.50\pm0.10$ & $-1.26\pm0.02$ & $-2.28\pm0.05$ & $68.0\pm2.0$ & $163\pm1.40$ & 2--400 & 4\\
041006 & 0.72 & $0.16\pm0.04$ & $-1.37\pm0.10$ & -- & $63.0\pm13.0$ & $12.0\pm1.00$ & 2--400 & 4\\
050318 & 1.44 & $0.21\pm0.07$* & $-0.34\pm0.32$ & -- & $47.0\pm15.0$ & $2.10\pm0.20$ & 15--350 & 4\\
050525 & 0.61 & $0.28\pm0.12$ & $-1.01\pm0.06$ & $-3.26\pm0.20$ & $81.2\pm1.4$ & $20.0\pm1.00$ & 15--350 & 4\\
051022 & 0.80 & $2.90\pm0.20$* & $-1.18\pm0.02$ & -- &  $510.0\pm22.0$ & $261\pm9.00$ & 20--2000 & 4\\ 
061121 & 1.31 & $2.31\pm1.16$ & $-1.32\pm0.05$ & -- & $606.0\pm90.0$ & $56.7\pm5.00$ & 20--5000 & 6, 7\\
061126 & 1.16 & $6.75\pm5.25$ & $-1.06\pm0.07$ & $-2.7\pm0.40$ & $620.0\pm220.0$ & $30.0\pm4.00$ & 15--10000 & 8,9\\ 
080319B &  0.94 & $0.03\pm0.01$ &  $-0.86\pm0.01$ &  $-3.59\pm0.62$ &  $675.0\pm22.0$ &  $613\pm13.0$ & 20--7000 & 
10\\
090328 & 0.74 & $6.40\pm6.0$ &  $-1.09\pm0.02$ &  $-2.37\pm0.18$ & $639.7\pm45.7$ & $50.9\pm0.82$ & 10--1000 & 
11,12\\
090618 & 0.54 & $0.60\pm0.05$ & $-1.13\pm0.01$ & $-2.22\pm0.02$ & $146.9\pm3.6$ & $257\pm1.50$ & 10--1000 & 13,12\\
091127 & 0.49 & $0.38\pm0.04$ & $-1.26\pm0.07$ & $-2.22\pm0.02$ & $35.5\pm1.5$ & $18.3\pm0.21$ & 10--1000 & 14,12\\
130427A & 0.34 & $0.43\pm0.40$ & $-0.91\pm0.01$ & $-3.18\pm0.03$ &  $877.8\pm4.9$ & $1900\pm2.0$ & 10--1000 & 15\\ 
\enddata
\\
\begin{flushleft}
The $\beta$ column contains the Band function high-energy power-law index.  The $\alpha$ column contains the Band 
function low-energy power-law index unless the corresponding $\beta$ values is missing.  In this case, $\alpha$ represents 
the spectral index from the Comptonized function.\\
$^\ast$ Either optical afterglow observations were unavailable or unable to constrain jet break time; X-ray break only.\\
$\dagger$ The first number references the jet break time source, and the second number references the spectral parameters 
source.  When there is just one number, the same source is used for the jet break time and the spectral parameters.
1-\citet{Bloom03}; 2-\citet{Goldstein13}; 3-\citet{Jimenez01}; 4-\citet{Schaefer07}; 5-\citet{Ghirlanda04}; 
6-\citet{Page07}; 7-\citet{Golenetskii06}; 8-\citet{Gomboc08}; 9-\citet{Perley08}; 10-\citet{Racusin08}; 11-\citet{Cenko11}; 12-
\citet{Goldstein12b}; 13-\citet{Page11}; 14-\citet{Filgas11}; 15-\citet{Maselli14}
\end{flushleft}
\hspace*{-2cm}
\end{deluxetable}

\begin{deluxetable}{ c  c  c  c  c  c  c  c  c }
\tablecolumns{9}
\tablewidth{0pt}
\tabletypesize{\scriptsize}
\setlength{\tabcolsep}{0.1in}
\tablecaption{GRB parameters  used to calculate jet opening angle  ($z>1.5$)\label{GRBHiZ}}
\startdata
\toprule
\rule[-2pt]{0pt}{12pt}	\multirow{2}{*}{GRB} & \multirow{2}{*}{$z$} & $t_j$ & \multirow{2}{*}{$\alpha$} & \multirow{2}{*}{$\beta$} & 
$E_{\rm peak}$ & Fluence & Band & \multirow{2}{*}{$\rm Ref.^{\dagger}$} \\  & & (days) &  & & (keV) & $(10^{-6}\ \rm erg \ 
cm^{-2})$ & (keV) & \\
\midrule
990123 & 1.60 & $2.04\pm0.46$ & $-0.90\pm0.10$ & $-2.48\pm0.40$ & $604.0\pm60.0$ & $270\pm30.0$ & 20--2000 & 1\\
990510 & 1.62 & $1.60\pm0.20$ & $-1.28\pm0.10$ & $-2.67\pm0.40$ & $126.0\pm10.0$ & $23.0\pm2.00$ & 20--2000 & 1\\
000926 & 2.04 & $1.80\pm0.10$ & $-1.10\pm0.10$ & $-2.43\pm0.40$ & $100.0\pm7.0$ & $6.20\pm0.60$ & 25--100 & 2,1\\
011211 & 2.14 & $1.56\pm0.16$ & $-0.84\pm0.09$ & -- & $59.0\pm8.0$ & $5.00\pm0.50$ & 40--700 & 1\\
020124 & 3.20 & $3.00\pm0.40$ & $-0.79\pm0.15$ & -- & $87.0\pm18.0$ & $8.10\pm0.80$ & 2--400 & 1\\
021004 & 2.33 & $4.74\pm0.50$ & $ -1.01\pm0.18$ &  -- & $80.0\pm53.0$ & $2.50\pm0.60$ & 2--400 & 1\\
030226 & 1.99 & $1.04\pm0.12$ & $-0.89\pm0.16$ & -- & $97.0\pm27.0$ & $5.60\pm0.70$ & 2--400 & 1\\
030328 & 1.52 & $0.80\pm0.10$ & $-0.80\pm0.80$ & $-2.30\pm0.00$ & $44.0\pm44.0$ & $0.65\pm0.28$ & 50--300 & 1,3\\
030429 & 2.66 & $1.77\pm1.00$ & $-1.12\pm0.24$ & -- & $35.0\pm12.0$ & $0.85\pm0.14$ & 2--400 & 1\\
050505 & 4.27 & $0.67\pm0.14$* & $-0.95\pm0.31$ & -- & $125.9\pm20.0$ & $15.8\pm0.16$ & 15--350 & 4,5\\
060124 & 2.30 & $1.10\pm0.10$* & $-1.29\pm0.07$ & $-2.25\pm0.30$ & $237.0\pm76.0$ & $28.0\pm3.00$ & 20--2000 & 1\\
060526 & 3.12 & $1.27\pm0.35$ & $-1.10\pm0.40$ & -- & $25.0\pm5.0$ & $0.49\pm0.06$ & 15--150 & 1\\
070125 & 1.55 & $3.80\pm0.10$ & $-1.13\pm0.09$ & $-2.08\pm0.14$ & $430.0\pm80.0$ & $179\pm13.0$ & 20--10000 & 6,7\\
090323 & 3.57 & $17.6\pm11.2$* & $-1.29\pm0.01$ & $-2.44\pm0.17$ & $632.9\pm40.8$ & $128\pm1.50$ & 10-1000 & 8,9\\
090902B & 1.82 & $6.20\pm0.80$ & $-1.01\pm0.01$ & -- & $1054\pm17.4$ & $266\pm0.77$ & 10--1000 & 8,9\\
090926A & 2.11 & $9.00\pm2.00$ & $-0.86\pm0.01$ & $-2.40\pm0.04$ & $340.0\pm5.7$ & $154\pm7.20$ & 10--1000 & 8,9\\
\enddata
\\
\begin{flushleft}
The $\beta$ column contains the Band function high-energy power-law index.  The $\alpha$ column contains 
the Band function low-energy power-law index unless the corresponding $\beta$ values is missing.  In this case, $\alpha$ 
represents the spectral index from the Comptonized function.\\
$^\ast$ Either optical afterglow observations were unavailable or unable to constrain jet break time; X-ray break only.\\
$\dagger$ The first number references the jet break time source, and the second number references the spectral parameters 
source.  When there is just one number, the same source is used for the jet break time and the spectral parameters.
1-\citet{Schaefer07}; 2-\citet{Bloom03}; 3-\citet{Atteia05}; 4-\citet{Hurkett06}; 5-\citet{Cabrera07}; 6-\citet{Chandra08}; 7-
\citet{Bellm08}; 8-\citet{Cenko11}; 9-\citet{Goldstein12b}
\end{flushleft}
\hspace*{-2cm}
\end{deluxetable}

\begin{deluxetable}{ c  c  c }
\tablecolumns{3}
\tablewidth{0pt}
\setlength{\tabcolsep}{0.2in}
\tablecaption{Log-normal distribution parameters. \label{distributionParams}}
\startdata
\toprule
Quantity & $\mu$ & $\sigma$\\
\midrule
$\theta_j$ & $0.77\pm0.02$ & $0.37\pm0.01$\\
$E_{iso}$ & $0.14\pm0.04$ & $0.84\pm0.03$\\
$E_\gamma$ & $-0.21\pm0.03$ & $0.64\pm0.03$\\
$L_{iso}$ & $-0.13\pm0.06$ & $0.96\pm0.05$\\
$L_\gamma$ & $0.38\pm0.06$ & $0.91\pm0.05$\\
$E_{p,z}$ & $2.69\pm0.02$ & $0.37\pm0.02$\\
$E_{p,z}$ (peak) & $2.80\pm0.02$ & $0.39\pm0.02$\\
$t_j$ & $0.86\pm0.05$ & $1.00\pm0.05$\\
\enddata
\end{deluxetable}

\begin{deluxetable}{ c  c  c }
\tablecolumns{3}
\tablewidth{0pt}
\setlength{\tabcolsep}{0.2in}
\tablecaption{Jet Angle and Energetics PDFs Table Format \label{pdfParams}}
\startdata
\toprule
Column & Format & Description\\
\midrule
1 & A9 & GBM Trigger \# \\
2 & F5.2 & $\theta_j$ log-normal mean \\
3 & F4.2 & $\theta_j$ log-normal std. dev.\\
4 & F5.2 & $E_{\rm iso}$ log-normal mean \\
5 & F4.2 & $E_{\rm iso}$ log-normal std. dev.\\
6 & F5.2 & $E_{\gamma}$ log-normal mean \\
7 & F4.2 & $E_{\gamma}$ log-normal std. dev.\\
8 & F5.2 & $L_{\rm iso}$ log-normal mean \\
9 & F4.2 & $L_{\rm iso}$ log-normal std. dev.\\
10 & F5.2 & $L_{\gamma}$ log-normal mean \\
11 & F4.2 & $L_{\gamma}$ log-normal std. dev.\\
12 & F4.2 & Time-integrated $E_{p,z}$ log-normal mean \\
13 & F4.2 &  Time-integrated $E_{p,z}$ log-normal std. dev.\\
14 & F4.2 & Peak $E_{p,z}$ log-normal mean \\
15 & F4.2 &  Peak $E_{p,z}$ log-normal std. dev.\\
14 & F5.2 & $t_j$ log-normal mean \\
15 & F4.2 &  $t_j$ log-normal std. dev.\\
\enddata
\end{deluxetable}

\end{document}